\pdfoutput=1
\documentclass[preprint,12pt]{elsarticle}

\usepackage[margin=1in]{geometry}

\usepackage[hyphens]{url}
\biboptions{sort&compress, square, comma}
\usepackage[breaklinks=true, linkcolor=blue, citecolor=blue, colorlinks=true]{hyperref}

\usepackage{graphicx}
\usepackage{caption}
\usepackage[font=footnotesize]{subcaption}

\usepackage[version=3]{mhchem} 
\usepackage{bm}
\usepackage{latexsym, amsmath,amssymb}

\usepackage{mathtools}
\usepackage{exscale, relsize}
\usepackage[retainorgcmds]{IEEEtrantools}

\usepackage{nicefrac}

\usepackage{booktabs,multicol}

\usepackage{cancel}
\usepackage[super]{nth}

\usepackage[T1]{fontenc}
\usepackage[english]{babel}
\usepackage{csquotes}
\usepackage{textcomp}

\usepackage{algorithm}
\usepackage[noend]{algpseudocode}
\usepackage{braket}
\DeclarePairedDelimiter\abs{\lvert}{\rvert}%
\DeclarePairedDelimiter\norm{\lVert}{\rVert}%

\MHInternalSyntaxOn
\renewcommand{\dcases}
 {
  \MT_start_cases:nnnn
    {\quad}
    {$\m@th\displaystyle##$\hfil}
    {$\m@th\displaystyle##$\hfil}
    {\lbrace}
 }
\MHInternalSyntaxOff

\makeatletter
\let\oldabs\abs
\def\abs{\@ifstar{\oldabs}{\oldabs*}}
\let\oldnorm\norm
\def\norm{\@ifstar{\oldnorm}{\oldnorm*}}
\makeatother

\usepackage{siunitx}
\DeclareSIUnit\atm{atm}

\def\CC{{C\nolinebreak[4]\hspace{-.05em}\raisebox{.4ex}{\footnotesize ++}}}
\newcommand{ \ddt } [1] { \frac{ \partial #1 }{ \partial t } }
\newcommand{ \ddx } [1] { \frac{ \partial }{ \partial #1 } }
\newcommand{ \dydx } [2] { \frac{ \partial #1 }{ \partial #2 } }

\newcommand{\pluseq}{\mathrel{{+}{=}}}

\newcommand{ \numsp }{N_{\text{sp}}}
\newcommand{ \numreac }{N_{\text{reac}}}

\graphicspath{{./figures/}}

\newcounter{bla}

\journal{Computer Physics Communications}

\begin{document}
\begin{frontmatter}

\title{\texttt{pyJac}: analytical Jacobian generator for chemical kinetics}

\author[osu]{Kyle~E.\ Niemeyer\corref{cor1}}
\ead{Kyle.Niemeyer@oregonstate.edu}

\author[uconn]{Nicholas~J.\ Curtis}
\author[uconn]{Chih-Jen Sung}

\address[osu]{School of Mechanical, Industrial, and Manufacturing Engineering\\
  Oregon State University, Corvallis, OR 97331, USA}
\address[uconn]{Department of Mechanical Engineering\\
  University of Connecticut, Storrs, CT, 06269, USA}

\cortext[cor1]{Corresponding author}

\begin{abstract}
Accurate simulations of combustion phenomena require the use of detailed chemical kinetics in order to capture limit phenomena such as ignition and extinction as well as predict pollutant formation.
However, the chemical kinetic models for hydrocarbon fuels of practical interest typically have large numbers of species and reactions and exhibit high levels of mathematical stiffness in the governing differential equations, particularly for larger fuel molecules.
In order to integrate the stiff equations governing chemical kinetics, generally reactive-flow simulations rely on implicit algorithms that require frequent Jacobian matrix evaluations.
Some in situ and a posteriori computational diagnostics methods also require accurate Jacobian matrices, including computational singular perturbation and chemical explosive mode analysis.
Typically, finite differences numerically approximate these, but for larger chemical kinetic models this poses significant computational demands since the number of chemical source term evaluations scales with the square of species count.
Furthermore, existing analytical Jacobian tools do not optimize evaluations or support emerging SIMD processors such as GPUs.
Here we introduce \texttt{pyJac}, a Python-based open-source program that generates analytical Jacobian matrices for use in chemical kinetics modeling and analysis.
In addition to producing the necessary customized source code for evaluating reaction rates (including all modern reaction rate formulations), the chemical source terms, and the Jacobian matrix, \texttt{pyJac} uses an optimized evaluation order to minimize computational and memory operations.
As a demonstration, we first establish the correctness of the Jacobian matrices for kinetic models of hydrogen, methane, ethylene, and isopentanol oxidation (number of species ranging \numrange{13}{360}) by showing agreement within \SI{0.001}{\percent} of matrices obtained via automatic differentiation.
We then demonstrate the performance achievable on CPUs and GPUs using \texttt{pyJac} via matrix evaluation timing comparisons; the routines produced by \texttt{pyJac} outperformed first-order finite differences by \numrange{3}{7.5} times and the existing analytical Jacobian software \texttt{TChem} by \numrange{1.1}{2.2} times on a single-threaded basis.
It is noted that \texttt{TChem} is not thread-safe, while \texttt{pyJac} is easily parallelized, and hence can greatly outperform \texttt{TChem} on multicore CPUs.
The Jacobian matrix generator we describe here will be useful for reducing the cost of integrating chemical source terms with implicit algorithms in particular and algorithms that require an accurate Jacobian matrix in general.
Furthermore, the open-source release of the program and Python-based implementation will enable wide adoption.
\end{abstract}


\end{frontmatter}



{\bf PROGRAM SUMMARY}

\begin{small}
\noindent
{\em Manuscript Title:} \texttt{pyJac}: analytical Jacobian generator for chemical kinetics \\
{\em Authors:} Kyle E.\ Niemeyer, Nicholas J.\ Curtis, Chih-Jen Sung \\
{\em Program Title:} pyJac \\
{\em Journal Reference:}                                      \\
{\em Catalogue identifier:}                                   \\
{\em Licensing provisions:} MIT License \\
{\em Programming language:} Python \\
{\em Computer:} Windows, Ubuntu, and Mac OS computers \\
{\em Operating system:} Any (Linux, Mac OS X, Windows) \\
{\em RAM:} Problem dependent; typically less than 1 GB \\
{\em Number of processors used:} one \\
{\em Classification:} 16.12, 4.3                                \\
{\em External routines/libraries:} required: \texttt{NumPy}; optional: \texttt{Cython}, Cantera, \texttt{PyYAML}, \texttt{Adept} \\
{\em Nature of problem:} Automatic generation of source code to evaluate Jacobian matrices for chemical kinetic models \\
{\em Solution method:} Chemical kinetic model interpreted from input file(s), and partial derivatives and necessary supporting functions are written to file based on a theoretical derivation.\\
{\em Additional comments:} This paper describes \texttt{pyJac} v1.0.2, available
at \url{http://dx.doi.org/10.5281/zenodo.251144}. Current version and support
available at \url{https://github.com/SLACKHA/pyjac/} \\
   \\
{\em Running time:} Problem dependent; from seconds to tens of minutes depending on the model size. 

\end{small}

\section{Introduction}
\label{sec:intro}

As the need for detailed and accurate chemical kinetic models\footnote{Note that the term ``reaction mechanism'' is also used commonly in the literature; we adopted our preferred terminology ``chemical kinetic model'' here.} in predictive reactive-flow simulations has become recognized in recent years, such models describing the oxidation of hydrocarbon fuels simultaneously grew orders of magnitude in size and complexity.
For example, a recently developed detailed kinetic model for 2-methylalkanes, relevant for jet and diesel fuel surrogates, consists of over 7000 species and 30000 reactions~\cite{Sarathy:2011kx}; similarly large surrogate models exist for gasoline~\cite{Mehl:2011cn,Mehl:2011jn} and biodiesel~\cite{Herbinet:2010gu}.
Since in general the computational cost of solving the associated generally stiff systems of equations scales quadratically with the number of species at best---and at worst, cubically---models of such a large size pose challenges even for lower-dimensional analyses, and cannot practically be used directly in multidimensional reactive-flow simulations.

In an effort to reduce the computational demands of using large, detailed kinetic models, a number of techniques have been developed to reduce their size and complexity while retaining predictiveness, as reviewed by Lu and Law~\cite{Lu:2009gh} as well as Tur{\'a}nyi and Tomlin~\cite{Turanyi:2014aa}.
Major classes of such approaches include skeletal reduction methods that remove unimportant species and reactions~\cite{Lu:2006bb,Pepiot-Desjardins:2008,Hiremath:2010jw,Niemeyer:2010bt}, lumping of species that share similar properties~\cite{Lu:2007,Ahmed:2007fa,Pepiot:2008kq}, and time-scale reduction methods that reduce chemical stiffness~\cite{Maas:1992ws,Lam:1994ws,Lu:2001ve,Gou:2010}.
Effective reduction strategies either combine multiple methods a priori~\cite{Lu:2008bi,Niemeyer:2014,Niemeyer:2015wq} or apply them dynamically during a simulation to achieve greater local savings~\cite{Banerjee:2006,Liang:2009,Shi:2010,Gou:2013eu,Yang:2013ip,Curtis:2015aa}.
Techniques such as interpolation\slash tabulation of expensive terms~\cite{Pope:1997wu} can also reduce computational costs.

In addition to the aforementioned cost reduction methods that modify the chemical kinetic models, improvements to the integration algorithms that actually solve the governing differential equations can also offer significant gains in computational performance.
Due to the chemical stiffness exhibited by most kinetic models, solvers typically rely on robust, high-order implicit integration algorithms based on backward differentiation formulae~\cite{Curtiss:1952,Byrne:1987wp,Brown:1989vl,Hindmarsh:2005hg}.
In order to solve the nonlinear algebraic equations that arise in these methods, the Jacobian matrix must be evaluated and factorized, operations that result in the quadratic and cubic costs mentioned previously.
However, by using an analytical formulation for the Jacobian matrix rather than a typical finite difference approximation, the cost of the numerous evaluations can drop from growing with the square of the number of species to a linear dependence~\cite{Lu:2009gh}.

In parallel with the potential improvements in stiff implicit integrators used for chemical kinetics, algorithms tailored for high-performance hardware accelerators offer another route to reducing computational costs.
In the past, central processing unit (CPU) clock speeds increased regularly---i.e., Moore's Law---but power consumption and heat dissipation issues disrupted this trend recently, slowing the pace of increases in CPU clock rates.
While multicore parallelism continues to raise CPU performance, single-instruction multiple data (SIMD) processors, e.g., graphics processing units (GPUs), recently emerged as a low-cost, low-power consumption, and massively parallel high-performance computing alternative.
GPUs---originally developed for graphics\slash video processing and display purposes---consist of hundreds to thousands of cores, compared to the tens of cores found on a typical CPU.
Recognizing that the SIMD parallelism model fits well with the operator-split chemistry integration that forms the basis of many reactive-flow codes~\cite{Oran:2001aa}, a number of studies in recent years~\cite{Spafford:2010aa,Shi:2011aa,Niemeyer:2011aa,Shi:2012aa,Stone:2013aa,Niemeyer:2014aa} explored the use of SIMD processors to accelerate the integration of chemical kinetics in reactive-flow codes.
Niemeyer and Sung~\cite{Niemeyer:2014ab} reviewed such efforts in greater detail.
While explicit methods offer significant improvements in performance for nonstiff and moderately stiff chemical kinetics~\cite{Niemeyer:2014aa}, experiences thus far suggest that stiff chemistry continues to require the use of implicit or similar algorithms.
This provides significant motivation to provide the capability of evaluating analytical Jacobian matrices on GPUs as well as CPUs.

Thus, motivated by the potential cost reductions offered by analytical Jacobian matrix formulations, over the past five years a number of research groups developed analytical Jacobian generators for chemical kinetics, although as will be discussed the software package introduced here offers a number of improvements.
The \texttt{TChem} toolkit developed by Safta et al.~\cite{Safta:2011vn} was one of the first software packages developed for calculating the analytical Jacobian matrix, but provides this functionality through an interface rather than generating customized source code for each model.
Youssefi~\cite{Youssefi:2011tm} recognized the importance of using an analytical Jacobian over a numerical approximation to reduce both computational cost and numerical error when performing eigendecomposition of the matrix.
Bisetti~\cite{Bisetti:2012jw} released a utility for producing analytical Jacobian matrix source code based on isothermal and isobaric conditions, with the state vector comprised of species concentrations rather than mass fractions; while incompatible with many existing reactive-flow formulations, this formulation resulted in a significant increase of Jacobian matrix sparsity---such a strategy should be investigated further.
Perini et al.~\cite{Perini:2012gy} developed an analytical Jacobian matrix formulation for constant-volume combustion, and when used in a multidimensional reactive-flow simulation---combined with tabulation of temperature-dependent properties---reported a performance improvement of around \SI{80}{\percent} over finite-difference-based approximations.
Recently, Dijkmans et al.~\cite{Dijkmans:2014bb} used a GPU-based analytical Jacobian combined with tabulation of temperature-dependent functions based on polynomial interpolations to accelerate the integration of chemical kinetics equations, similar to the earlier approach of Shi et al.~\cite{Shi:2011aa}.
Unlike the current work, the approach of Dijkmans et al.~\cite{Dijkmans:2014bb} used the GPU to calculate the elements of a single Jacobian matrix in parallel, rather than a large number of matrices corresponding to different states.

To our knowledge, currently no open-source analytical chemical Jacobian tool exists that is capable of generating code specifically optimized for SIMD processors.
To this end, \texttt{pyJac} is capable of generating subroutines for reaction rates, species production rates, derivative source terms, and analytical chemical Jacobian matrices both for CPU operation via C\slash \CC{} and GPU operation via CUDA~\cite{Nickolls:2008aa}, a widely used programming language for NVIDIA GPUs.
Furthermore, \texttt{pyJac} supports newer pressure-dependent reaction formulations (i.e., based on logarithmic or Chebyshev polynomial interpolation).

The rest of the paper is structured as follows.
First, in Section~\ref{S:theory} we introduce the governing equations for chemical kinetics and then provide the analytical Jacobian matrix formulation in Section~\ref{S:jacobian}.
Next, Section~\ref{S:opt} describes the techniques used to optimize evaluation of the analytical Jacobian on both CPUs and GPUs.
Then, in Section~\ref{S:results} we demonstrate the correctness and computational performance of the generated analytical Jacobian matrices for benchmark chemical kinetic models, and discuss the implications of these results.
Finally, we summarize our work in Section~\ref{S:conclusions} and outline future research directions.

\section{Theory}
\label{S:theory}

This section describes the theoretical background of the analytical Jacobian generator, first in terms of the governing equations and then the components of the Jacobian matrix itself.
The literature contains more detailed explanations of the governing equations development~\cite{Law:2006tu,Warnatz:2006tq,Glassman:2008tq}, but we include the necessary details here for completeness.

\subsection{Governing equations}
\label{sec:goveq}

The initial value problem to be solved, whether in the context of a single homogeneous reacting system (e.g., autoignition, perfectly stirred reactor) or the chemistry portion of an operator-split multidimensional reactive-flow simulation~\cite{Oran:2001aa}, is described using an ordinary differential equation for the thermochemical composition vector
\begin{equation}
\label{e:vars}
\Phi = \left \lbrace T, Y_1, Y_2, \dotsc, Y_{\numsp - 1} \right \rbrace^{\intercal} \;,
\end{equation}
where $T$ is the temperature, $Y_i$ are the species mass fractions, and $\numsp$ is the number of species.
The mass fraction of the final species, $Y_{\numsp}$, is determined through conservation of mass:
\begin{equation}
Y_{\numsp} = 1 - \sum_{k=1}^{\numsp - 1} Y_k \;,
\label{e:y_nsp}
\end{equation}
where the most abundant species (e.g., \ce{N2} in air-fed combustion) can be assigned to this role.
In multidimensional simulations where the equations for chemical kinetics are coupled to conservation of energy (or enthalpy), temperature can be determined algebraically in a straightforward manner~\cite{Oran:2001aa}.
Completely defining the thermodynamic state also requires either pressure ($p$) or density ($\rho$), related to temperature and the mixture composition through the ideal equation of state
\begin{equation}
\label{e:state}
p = \rho \frac{\mathcal{R}}{W} T = \mathcal{R} T \sum_{k=1}^{\numsp} [X_k] \;,
\end{equation}
where $\mathcal{R}$ is the universal gas constant, $W$ is the average molecular weight of the mixture, and $[X_k]$ is the molar concentration of the $k$th species.
In the current implementation of \texttt{pyJac}, we assume a constant-pressure system\footnote{In the context of multidimensional reactive flows, the constant-pressure assumption signifies that pressure remains constant during the reaction substep of an operator-split scheme (and is thus an input variable to the kinetics problem), not that the pressure is fixed throughout the entire simulation. This assumption is commonly used in low-speed combustion simulations.}; thus, we use Eq.~\eqref{e:state} to determine density rather than including it in the differential system given by Eq.~\eqref{e:vars}.
The average molecular weight is defined by
\begin{equation}
W = \frac{1}{\sum_{k=1}^{\numsp} Y_k / W_k} = \frac{\rho \mathcal{R} T}{p}
\end{equation}
and the molar concentrations by
\begin{equation}
[X_k] = \rho \frac{Y_k}{W_k} \;,
\end{equation}
where $W_k$ is the molecular weight of the $k$th species.

The system of ODEs governing the time change in thermochemical composition corresponding to Eq.~\eqref{e:vars} is then $ f = \partial \Phi/ \partial t$:
\begin{equation}
f = \ddt{\Phi} = \left \lbrace \ddt{T}, \ddt{Y_1}, \ddt{Y_2}, \dotsc, \ddt{Y_{\numsp - 1}} \right\rbrace^{\intercal} \;,
\label{e:ode}
\end{equation}
where
\begin{align}
\ddt{T} &= \frac{-1}{\rho c_p} \sum_{k=1}^{\numsp} h_k W_k \dot{\omega}_k \;, \\
\ddt{Y_k} &= \frac{W_k}{\rho} \dot{\omega}_k \quad k = 1, \dotsc, \numsp - 1 \;, \label{e:dTdt}
\end{align}
where $c_p$ is the mass-averaged constant-pressure specific heat, $h_k$ is the enthalpy of the $k$th species in mass units, and $\dot{\omega}_k$ is the $k$th species overall production rate.

\subsection{Thermodynamic properties}

The standard-state thermodynamic properties (in molar units) for a gaseous species $k$ is defined using the standard seven-coefficient polynomial of Gordon and McBride~\cite{Gordon:1976wp}:
\begin{align}
\frac{C_{p,k}^{\circ}}{\mathcal{R}} &= a_{0,k} + T \left( a_{1,k} + T \left( a_{2,k} + T \left( a_{3,k} + a_{4,k} T \right) \right) \right) \label{e:cpk} \\
\frac{H_k^{\circ}}{\mathcal{R}} &= T \left( a_{0,k} + T \left( \frac{a_{1,k}}{2} + T \left( \frac{a_{2,k}}{3} + T \left( \frac{a_{3,k}}{4} + \frac{a_{4,k}}{5} T \right) \right) \right) \right) + a_{5,k} \label{e:hk} \\
\frac{S_k^{\circ}}{\mathcal{R}} &= a_{0,k} \ln T + T \left( a_{1,k} + T \left( \frac{a_{2,k}}{2} + T \left( \frac{a_{3,k}}{3} + \frac{a_{4,k}}{4} T \right) \right) \right) + a_{6,k} \label{e:sk}
\end{align}
where $C_{p,k}$ is the constant-pressure specific heat in molar units, $H_k$ is the enthalpy in molar units, $S_k$ is the entropy in molar units, and the ${}^{\circ}$ indicates a standard-state property at one atmosphere (equivalent to the property at any pressure for calorically perfect gases).

The mass-based specific heat and enthalpy are then defined as
\begin{equation}
c_{p,k} = \frac{C_{p,k}}{W_k} \quad \text{and} \quad h_k = \frac{H_k}{W_k} \;,
\end{equation}
and the mixture-averaged specific heat is
\begin{equation}
c_p = \sum_{k=1}^{\numsp} Y_k c_{p,k} \;.
\end{equation}

\subsection{Reaction rate expressions}

Next, we define the species rates of production and related kinetic terms as
\begin{equation}
\dot{\omega}_k = \sum_{i=1}^{N_{\text{reac}}} \nu_{k i} q_i \;,
\end{equation}
where $N_{\text{reac}}$ is the number of reactions, $\nu_{k i}$ is the overall stoichiometric coefficient for species $k$ in reaction $i$, and $q_i$ is the rate-of-progress for reaction $i$.
These are defined by
\begin{align}
\nu_{k i} &= \nu_{k i}^{\prime \prime} - \nu_{k i}^{\prime}  \quad \text{and} \\
q_i &= c_i R_i \;,
\end{align}
where $\nu_{k i}^{\prime \prime}$ and $\nu_{k i}^{\prime}$ are the product and reactant stoichiometric coefficients (respectively) of species $k$ in reaction $i$.
The base rate-of-progress for the $i$th reversible reaction $R_i$ is given by
\begin{align}
R_i &= R_{f, i} - R_{r, i} \;, \\
R_{f, i} &= k_{f, i} \prod_{j = 1}^{\numsp} [X_j]^{\nu_{j i}^{\prime}} \;, \text{ and} \\
R_{r, i} &= k_{r, i} \prod_{j = 1}^{\numsp} [X_j]^{\nu_{j i}^{\prime \prime}} \;,
\end{align}
where $k_{f,i}$ and $k_{r,i}$ are the forward and reverse reaction rate coefficients for the $i$th reaction, respectively, and the third-body\slash pressure modification $c_i$ is given by
\begin{equation}
c_i = \begin{dcases}
  1 &\text{for elementary reactions,} \\
  [X]_i &\text{for third-body enhanced reactions,} \\
  \frac{P_{r,i}}{1 + P_{r,i}} F_i &\text{for unimolecular/recombination falloff reactions, and} \\
  \frac{1}{1 + P_{r,i}} F_i &\text{for chemically-activated bimolecular reactions,}
  \end{dcases}
\label{e:rxn_pressure}
\end{equation}
where for the $i$th reaction $[X]_i$ is the third-body concentration, $P_{r,i}$ is the reduced pressure, and $F_i$ is the falloff blending factor.
These terms are defined in the following sections.

The forward reaction rate coefficient $k_{f, i}$ is given by the three-parameter Arrhenius expression:
\begin{equation}
\label{e:arrhenius}
  k_{f, i} = A_i T^{\beta_i} \exp \left( - \frac{T_{a, i}}{T} \right) \;,
\end{equation}
where $A_i$ is the pre-exponential factor, $\beta_i$ is the temperature exponent, and $T_{a, i}$ is the activation temperature given by $T_{a, i} = E_{a, i} / \mathcal{R}$.

As given by Lu and Law~\cite{Lu:2009gh}, depending on the value of the Arrhenius parameters, $k_{f,i}$ can be calculated in different ways to minimize the computational cost:
\begin{equation}
\label{e:kf_cost}
  k_{f,i} =
  \begin{dcases}
  A_i & \text{if } \beta = 0 \text{ and } T_{a,i} = 0 \;, \\
  \exp \left( \log A_i + \beta_i \log T \right)   & \text{if } \beta_i \neq 0 \text{ and } T_{a, i} = 0 \;, \\
  \exp \left( \log A_i + \beta_i \log T - T_{a, i} / T \right) & \text{if } \beta_i \neq 0 \text{ and } T_{a, i} \neq 0 \;, \\
  \exp \left( \log A_i - T_{a, i} / T \right)  & \text{if } \beta_i = 0 \text{ and } T_{a, i} \neq 0 \;, \text{ and} \\
  A_i \prod^{\beta_i} T & \text{if } T_{a, i} = 0 \text{ and } \beta_i \in \mathbb{Z} \;,
  \end{dcases}
\end{equation}
where $\mathbb{Z}$ is the set of integers.

\subsubsection{Reverse rate coefficient}

By definition, irreversible reactions have a zero reverse rate coefficient $k_{r, i}$, while reversible reactions have nonzero $k_{r, i}$.
For reversible reactions, the reverse rate coefficient is determined in one of two ways: (1) via explicit reverse Arrhenius parameters as with the forward rate coefficient---thus following the same expression as Eq.~\eqref{e:arrhenius}---or (2) via the ratio of the forward rate coefficient and the equilibrium constant,
{\allowdisplaybreaks \begin{IEEEeqnarray}{rCl}
k_{r, i} &=& \frac{k_{f, i}}{K_{c, i}} \;, \label{e:kri} \\
K_{c, i} &=& K_{p, i} \left( \frac{p_{\text{atm}}}{\mathcal{R} T} \right)^{\sum_{k=1}^{\numsp} \nu_{k i}} \;, \text{ and} \\
K_{p, i} &=& \exp \left( \frac{\Delta S_i^{\circ}}{\mathcal{R}} - \frac{\Delta H_i^{\circ}}{\mathcal{R} T} \right) = \exp \left( \sum_{k=1}^{\numsp} \nu_{k i} \left( \frac{S_k^{\circ}}{\mathcal{R}} - \frac{H_k^{\circ}}{\mathcal{R} T} \right) \right) \;,
\end{IEEEeqnarray}}%
where $p_{\text{atm}}$ is the pressure of one standard atmosphere in the appropriate units.

By combining the expressions for $K_{c, i}$ and $K_{p, i}$, we obtain
\begin{align}
K_{c, i}
 &= \left( \frac{p_{\text{atm}}}{\mathcal{R}} \right)^{\sum_{k=1}^{\numsp} \nu_{k i}} \exp \left( \sum_{k=1}^{\numsp} \nu_{k i} B_k \right) \;,
\end{align}
where, expanding the polynomial expressions for $S_k^{\circ}$ and $H_k^{\circ}$ from Eqs.~\eqref{e:sk} and \eqref{e:hk}, respectively,
{\allowdisplaybreaks \begin{IEEEeqnarray}{rCl}
B_k
  & = & a_{6,k} - a_{0,k} + \left( a_{0,k} - 1 \right) \ln T \nonumber \\
  & & +\: T \left( \frac{a_{1,k}}{2} + T \left( \frac{a_{2,k}}{6} + T \left( \frac{a_{3,k}}{12} + \frac{a_{4,k}}{20} T \right) \right) \right) - \frac{a_{5,k}}{T} \;. \IEEEeqnarraynumspace
\end{IEEEeqnarray}}%

\subsubsection{Third-body effects}

For a reaction enhanced (or diminished) by the presence of a third body, the reaction rate is modified by the third-body concentration $[X]_i$ given by
\begin{equation}
[X]_i = \sum_{j=1}^{\numsp} \alpha_{i j} [X_j] \;,
\end{equation}
where $\alpha_{i j}$ is the third-body efficiency of species $j$ in the $i$th reaction.
If all species in the mixture contribute equally as third bodies (the default) then $\alpha_{i j} = 1$ for all species.
In this case,
\begin{equation}
[X]_i = [M] = \sum_{j=1}^{\numsp} [X_j] = \frac{p}{\mathcal{R} T} = \frac{\rho}{W} \;.
\end{equation}
In addition, a single species may act as the third body, in which case
\begin{equation}
[X]_i = [X_m] \;,
\end{equation}
where the $m$th species is the third body.

\subsubsection{Falloff reactions}

Unlike elementary and third-body reactions, falloff reactions exhibit a pressure dependence described as a blending of rates at low- and high-pressure limits; thus, the rate coefficients depend on a mixture of low-pressure- ($k_{0, i}$) and high-pressure-limit ($k_{\infty,i}$) coefficients, each with corresponding Arrhenius parameters and expressed using Eq.~\eqref{e:arrhenius}.
The ratio of the coefficients $k_{0, i}$ and $k_{\infty, i}$, combined with the third-body concentration (based on either the mixture as a whole including any efficiencies $\alpha_{i,j}$, or a specific species), define a reduced pressure $P_{r,i}$ given by
\begin{equation}
P_{r,i} = \begin{dcases}
\frac{k_{0,i}}{k_{\infty,i}} [X]_i &\text{for the mixture as the third body, or} \\
\frac{k_{0,i}}{k_{\infty,i}} [X_m] &\text{for a specific species } m \text{ as third body.}
\end{dcases}
\label{e:pr_i}
\end{equation}

The falloff blending factor $F_i$ used in Eq.~\eqref{e:rxn_pressure} is determined based on one of three representations: the Lindemann~\cite{Lindemann:1922cz}, Troe~\cite{Gilbert:1983bb}, and SRI~\cite{Stewart:1989gj} falloff approaches
\begin{equation}
F_i = \begin{dcases}
1 &\text{for Lindemann,} \\
F_{\text{cent}}^{ \left( 1 + ( A_{\text{Troe}} / B_{\text{Troe}} )^2 \right)^{-1} } &\text{for Troe, or} \\
d T^e \left( a \cdot \exp \left( -\frac{b}{T} \right) + \exp \left( -\frac{T}{c} \right) \right)^X &\text{for SRI.}
\end{dcases}
\end{equation}

The variables for the Troe representation are given by
\begin{align}
F_{\text{cent}} &= (1 - a) \exp \left( -\frac{T}{T^{***}} \right) + a\cdot \exp \left( -\frac{T}{T^*} \right) + \exp \left( -\frac{T^{**}}{T} \right) \;, \\
A_{\text{Troe}} &= \log_{10} P_{r,i} - 0.67 \log_{10} F_{\text{cent}} - 0.4 \;, \text{and} \\
B_{\text{Troe}} &= 0.806 - 1.1762 \log_{10} F_{\text{cent}} - 0.14 \log_{10} P_{r,i} \;,
\end{align}
where $a$, $T^{***}$, $T^*$, and $T^{**}$ are specified parameters.
The final parameter $T^{**}$ is optional, and, if it is not used, the final term of $F_{\text{cent}}$ is omitted.
The exponent used in the SRI representation is given by
\begin{equation}
X = \left( 1 + \left( \log_{10} P_{r,i} \right)^2 \right)^{-1}
\end{equation}
where $a$, $b$, and $c$ are required parameters.
The parameters $d$ and $e$ are optional; if not specified, $d = 1$ and $e = 0$.

\subsubsection{Pressure-dependent reactions}

In addition to the falloff approach given previously, two additional formulations can be used to describe the pressure dependence of reactions that do not follow the modification factor $c_i$ approach.
The first involves logarithmic interpolation between Arrhenius rates at two pressures~\cite{chemkin:2012,Goodwin:2015aa}, each evaluated using Eq.~\eqref{e:arrhenius}:
\begin{align}
k_1 (T) &= A_1 T^{\beta_1} \exp \left( -\frac{T_{a, 1}}{T} \right) \text{ at } p_1 \text{ and} \label{e:plog_k1} \\
k_2 (T) &= A_2 T^{\beta_2} \exp \left( -\frac{T_{a, 2}}{T} \right) \text{ at } p_2 \;, \label{e:plog_k2}
\end{align}
where the Arrhenius coefficients are given for each pressure $p_1$ and $p_2$.
Then, the reaction rate coefficient at a particular pressure $p$ between $p_1$ and $p_2$ can be determined through logarithmic interpolation:
\begin{equation}
\log k_f (T, p) = \log k_1 (T) + \left( \log k_2 (T) - \log k_1 (T) \right) \frac{\log p - \log p_1}{\log p_2 - \log p_1} \;. \label{e:plog}
\end{equation}

In addition, the pressure dependence of a reaction can be expressed through a bivariate Chebyshev polynomial fit~\cite{Venkatesh:1997hv,Venkatesh:1997ik,Venkatesh:2000gj,chemkin:2012,Goodwin:2015aa}:
\begin{equation}
\log_{10} k_f (T, p) = \sum_{i = 1}^{N_T} \sum_{j = 1}^{N_p} \eta_{ij} \phi_i (\tilde{T}) \phi_j \left(\tilde{p}\right) \label{e:cheb} \;,
\end{equation}
where $\eta_{ij}$ is the coefficient corresponding to the grid points $i$ and $j$, $N_T$ and $N_p$ are the numbers of grid points for temperature and pressure, respectively, and $\phi_n$ is the Chebyshev polynomial of the first kind of degree $n - 1$ typically expressed as
\begin{equation}
\phi_n (x) = \mathcal{T}_{n-1} (x) = \cos \left( (n - 1) \arccos (x) \right) \quad \text{for } |x| \leq 1 \;.
\end{equation}
The reduced temperature $\tilde{T}$ and pressure $\tilde{p}$ are given by
\begin{align}
\tilde{T} &\equiv \frac{2 T^{-1} - T^{-1}_{\min} - T^{-1}_{\max}}{T^{-1}_{\max} - T^{-1}_{\min}} \quad\text{and} \\
\tilde{p} &\equiv \frac{2\log_{10} p - \log_{10} p_{\min} - \log_{10} p_{\max}}{\log_{10} p_{\max} - \log_{10} p_{\min}} \;,
\end{align}
where $T_{\min} \leq T \leq T_{\max}$ and $p_{\min} \leq p \leq p_{\max}$ describe the ranges of validity for temperature and pressure.

\section{Jacobian matrix formulation}
\label{S:jacobian}

Next, we detail the construction of the Jacobian matrix, and provide a simple method for evaluating it efficiently by calculating elements in the appropriate order.
More sophisticated approaches for reducing the cost of evaluating the matrix will be proposed in Section~\ref{S:opt}.

\subsection{Elements of the Jacobian matrix}

Let $\mathcal{J}$ denote the Jacobian matrix corresponding to the vector of ODEs given by Eq.~\eqref{e:ode}.
$\mathcal{J}$ is filled by the partial derivatives $\partial f / \partial \Phi$, such that
\begin{equation}
\mathcal{J}_{i,j} = \dydx{f_i}{\Phi_j} \;.
\end{equation}


The first line of $\mathcal{J}$ is filled with partial derivatives of the energy equation, or
\begin{equation}
\mathcal{J}_{1,j} = \dydx{\dot{T}}{\Phi_j} \quad j = 1, \dotsc, \numsp - 1 \;.
\end{equation}
The components of $\mathcal{J}_{1,j}$ are:
{\allowdisplaybreaks \begin{IEEEeqnarray}{rCl}
\mathcal{J}_{1,1} &=& \dydx{f_1}{T}
= \frac{-1}{c_p} \sum_{k=1}^{\numsp} \left[ \left( \dydx{h_k}{T} - h_k \dydx{c_p}{T} \right) \frac{W_k \dot{\omega}_k}{\rho} + \frac{h_k}{c_p} \ddx{T} \left( \frac{W_k \dot{\omega}_k}{\rho} \right) \right] , \IEEEeqnarraynumspace \\
\mathcal{J}_{1, j+1} &=& \dydx{f_1}{Y_j}
= \frac{-1}{c_p} \sum_{k=1}^{\numsp} \left[ \left( \dydx{h_k}{Y_j} - h_k \dydx{c_p}{Y_j} \right) \frac{W_k \dot{\omega}_k}{\rho} + \frac{h_k}{c_p} \ddx{Y_j} \left( \frac{W_k \dot{\omega}_k}{\rho} \right) \right] , \IEEEeqnarraynumspace
\end{IEEEeqnarray}}%
for $j = 1, \dotsc, \numsp - 1$.


The remaining lines of $\mathcal{J}$ are filled with the partial derivatives of the species equations, with the components
{\allowdisplaybreaks \begin{IEEEeqnarray}{rCl}
\mathcal{J}_{k+1, 1} & = & \dydx{f_{k+1}}{T}
 =  \frac{W_k}{\rho} \left(\dydx{\dot{\omega}_k}{T} - \frac{\dot{\omega}_k}{\rho} \dydx{\rho}{T} \right) \text{ and} \IEEEeqnarraynumspace \\
\mathcal{J}_{k+1, j+1} & = & \dydx{f_{k+1}}{Y_j}
= \frac{W_k}{\rho} \left( \dydx{\dot{\omega}_k}{Y_j} - \frac{\dot{\omega}_k}{\rho} \dydx{\rho}{Y_j} \right) , \IEEEeqnarraynumspace
\label{e:jac_spec}
\end{IEEEeqnarray}}%
for $k = 1, \dotsc, \numsp - 1$ and $j = 1, \dotsc, \numsp - 1$.


The partial derivatives of $\rho$, $c_p$, $h_k$, and $\dot{\omega}_k$ with respect to temperature and mass fraction need to be evaluated.
The partial derivatives of density are
{\allowdisplaybreaks \begin{IEEEeqnarray}{rCl}
\dydx{\rho}{T} & = & \ddx{T} \left( \frac{p W}{\mathcal{R} T} \right)
 =  -\frac{\rho}{T} \quad \text{and} \\
\dydx{\rho}{Y_j} & = & \ddx{Y_j} \left( \frac{p W}{\mathcal{R} T} \right)
= -\rho W \left( \frac{1}{W_j} - \frac{1}{W_{\numsp}} \right) \;.
\end{IEEEeqnarray}}%
The partial derivative of species specific enthalpy with respect to temperature is simply constant-pressure specific heat, and the partial derivative with respect to species mass fraction is zero:
\begin{align}
\dydx{h_k}{T} &= c_{p,k} \\
\dydx{h_k}{Y_j} &= 0 \;.
\end{align}
The derivatives of specific heat are expressed by
\begin{align}
\dydx{c_p}{T} &= \sum_{k=1}^{\numsp} Y_k \dydx{c_{p,k}}{T} \;, \label{E:dcpdT} \\
\dydx{c_{p,k}}{T} &= \frac{\mathcal{R}}{W_k} \left( a_{1,k} + T \left( 2 a_{2,k} + T \left( 3 a_{3,k} + 4 a_{4,k} T \right) \right) \right) \;, \text{ and} \\
\dydx{c_p}{Y_j} &= \ddx{Y_j} \sum_{k=1}^{\numsp} Y_k c_{p,k} = c_{p,j} - c_{p, \numsp} \;.
\end{align}

Next, the derivatives of species rate-of-production $\dot{\omega}_k$ are given by
{\allowdisplaybreaks \begin{IEEEeqnarray}{rCl}
\dydx{\dot{\omega}_k}{T} & = & \sum_{i=1}^{N_{\text{reac}}} \nu_{k i} \dydx{q_i}{T} = \sum_{i=1}^{N_{\text{reac}}} \nu_{ki} \left( \dydx{c_i}{T} R_i + c_i \dydx{R_i}{T} \right) \text{ and} \\
\dydx{\dot{\omega}_k}{Y_j} &=& \sum_{i=1}^{N_{\text{reac}}} \nu_{ki} \dydx{q_i}{Y_j} = \sum_{i=1}^{N_{\text{reac}}} \nu_{ki} \left( \dydx{c_i}{Y_j} R_i + c_i \dydx{R_i}{Y_j} \right) \;.
\end{IEEEeqnarray}}%
The partial derivatives of $R_i$ vary depending on whether the reaction is reversible.
For irreversible reactions,
{\allowdisplaybreaks \begin{IEEEeqnarray}{rCl}
\dydx{R_{f, i}}{T}
&=& \dydx{k_{f, i}}{T} \frac{1}{k_{f, i}} R_{f, i} + k_{f, i} \ddx{T} \left( \prod_{j=1}^{\numsp} [X_j]^{\nu_{ji}^{\prime}} \right) \; \text{and} \label{E:dRfdT-orig} \\
\dydx{R_{f, i}}{Y_j} &=& k_{f, i} \ddx{Y_j} \left( \prod_{k=1}^{\numsp} [X_k]^{\nu_{ki}^{\prime}} \right) \;. \label{E:dRfdY-orig}
\end{IEEEeqnarray}}%
The partial derivatives of species concentration are
{\allowdisplaybreaks \begin{IEEEeqnarray}{rCl}
\dydx{[X_i]}{T} &=& \frac{Y_i}{W_i} \dydx{\rho}{T}
= -\frac{[X_i]}{T} \\
\dydx{[X_i]}{Y_j} &=& \frac{Y_i}{W_i} \dydx{\rho}{Y_j} + \frac{\rho}{W_i} \dydx{Y_i}{Y_j}
= -[X_i] W \left(\frac{1}{W_j} - \frac{1}{W_{\numsp}} \right) + \left( \delta_{i j} - \delta_{i \numsp} \right) \frac{\rho}{W_i} \;,
\end{IEEEeqnarray}}%
where $i = 1, \ldots, \numsp$, $j = 1, \ldots, \numsp - 1$, and $\delta_{ij}$ is the Kronecker delta.
The partial derivatives of the molar concentration product terms are then
{\allowdisplaybreaks \begin{IEEEeqnarray}{rCl}
\ddx{T} \left( \prod_{j=1}^{\numsp} [X_j]^{\nu_{ji}^{\prime}} \right)
&=& -\frac{1}{T} \left( \prod_{j=1}^{\numsp} [X_j]^{\nu_{ji}^{\prime}} \right) \sum_{j=1}^{\numsp} \nu_{ji}^{\prime} \quad \text{ and} \\
\ddx{Y_j} \left( \prod_{k=1}^{\numsp} [X_k]^{\nu_{ki}^{\prime}} \right)
&=& \sum_{k=1}^{\numsp} \nu_{ki}^{\prime} \Biggl( -[X_k]^{\nu_{ki}^{\prime}} W \left(\frac{1}{W_j} - \frac{1}{W_{\numsp}} \right) \nonumber \\
&& +\: \left( \delta_{k j} - \delta_{k \numsp} \right) \frac{\rho}{W_k} [X_k]^{\nu_{ki}^{\prime} - 1} \Biggr) \prod_{\substack{l=1 \\ l \ne k}}^{\numsp} [X_l]^{\nu_{li}^{\prime}} \;,
\end{IEEEeqnarray}}%
and the partial derivative of forward reaction rate coefficient is
{\allowdisplaybreaks \begin{IEEEeqnarray}{rCl}
  \dydx{k_{f, i}}{T}
  & = & \frac{k_{f,i}}{T} \left( \beta_i + \frac{T_{a,i}}{T} \right) \;.
\label{e:dkfdT}
\end{IEEEeqnarray}}%
Inserting these into Eqs.~\eqref{E:dRfdT-orig} and \eqref{E:dRfdY-orig} gives
{\allowdisplaybreaks \begin{IEEEeqnarray}{rCl}
\dydx{R_{f, i}}{T}
&=& \frac{R_{f,i}}{T} \left( \beta_i + \frac{T_{a,i}}{T} - \sum_{j=1}^{\numsp} \nu_{ji}^{\prime} \right) \text{ and} \label{E:dRfdT} \\
\dydx{R_{f,i}}{Y_j} &=& \sum_{k=1}^{\numsp} \nu_{ki}^{\prime} \Biggl( -R_{f,i} W \left(\frac{1}{W_j} - \frac{1}{W_{\numsp}} \right) \nonumber \\
&& +\: \left( \delta_{k j} - \delta_{k \numsp} \right) k_{f,i} \frac{\rho}{W_k} [X_k]^{\nu_{ki}^{\prime} - 1} \prod_{\substack{l=1 \\ l \ne k}}^{\numsp} [X_l]^{\nu_{li}^{\prime}} \Biggr) \;. \IEEEeqnarraynumspace \label{E:dRfdY}
\end{IEEEeqnarray}}%

For reversible reactions with explicit reverse Arrhenius coefficients,
{\allowdisplaybreaks \begin{IEEEeqnarray}{rCl}
\dydx{R_i}{T}
& = & \dydx{k_{f, i}}{T} \prod_{j=1}^{\numsp} [X_j]^{\nu_{ji}^{\prime}} + k_{f, i} \ddx{T} \left( \prod_{j=1}^{\numsp} [X_j]^{\nu_{ji}^{\prime}} \right) \nonumber \\
&& -\: \dydx{k_{r, i}}{T} \prod_{j=1}^{\numsp} [X_j]^{\nu_{ji}^{\prime \prime}} - k_{r, i} \ddx{T} \left( \prod_{j=1}^{\numsp} [X_j]^{\nu_{ji}^{\prime\prime}} \right) \label{e:dRdT_explicit_rev_general} \\
&=& \frac{R_{f,i}}{T} \left( \beta_{f,i} + \frac{T_{af,i}}{T} - \sum_{j=1}^{\numsp} \nu_{ji}^{\prime} \right) - \frac{R_{r,i}}{T} \left( \beta_{r,i} + \frac{T_{ar,i}}{T} - \sum_{j=1}^{\numsp} \nu_{ji}^{\prime\prime} \right) \;. \IEEEeqnarraynumspace
\end{IEEEeqnarray}}%
Similarly, $ \frac{1}{ [X_j] } \left( \nu_{ji}^{\prime} R_{f, i} - \nu_{ji}^{\prime \prime} R_{r, i} \right) $ should be avoided since $ [X_j] $ could be zero.

The partial derivative of the reverse reaction rate coefficient, when evaluated using the forward rate coefficient and equilibrium constant using Eq.~\eqref{e:kri}, is more complicated:
{\allowdisplaybreaks \begin{IEEEeqnarray}{rCl}
  \dydx{k_{r, i}}{T}
  & = & \ddx{T} \left( \frac{k_{f, i}}{K_{c, i}} \right) = \frac{ \dydx{k_{f, i}}{T} K_{c, i} - \dydx{K_{c, i}}{T} k_{f, i} }{K_{c, i}^2} \label{e:dkrdT_general} \\
  & = & \frac{k_{f, i}}{K_{c, i}} \frac{1}{T} \left( \beta_i + \frac{T_{a, i}}{T} \right) - \frac{1}{K_{c, i}} \dydx{K_{c, i}}{T} \frac{k_{f, i}}{K_{c, i}} \\
  \dydx{K_{c, i}}{T}
  & = & K_{c, i} \sum_{k=1}^{\numsp} \nu_{ki} \dydx{B_k}{T} \\
  \therefore \dydx{k_{r, i}}{T} & = & k_{r, i} \left( \frac{1}{T} \left( \beta_i + \frac{T_{a, i}}{T} \right) - \sum_{k=1}^{\numsp} \nu_{ki} \dydx{B_k}{T} \right) \;,
\end{IEEEeqnarray}}%
where
\begin{equation}
  \dydx{B_k}{T} = \frac{1}{T} \left( a_{0,k} - 1 + \frac{a_{5,k}}{T} \right) + \frac{a_{1,k}}{2} + T \biggl( \frac{a_{2,k}}{3} + T \left( \frac{a_{3,k}}{4} + \frac{a_{4,k}}{5} T \right) \biggr) \;.
\end{equation}
Now, the partial derivative of $R_i$ with respect to temperature is
{\allowdisplaybreaks \begin{IEEEeqnarray}{rCl}
\dydx{R_i}{T}
&=& \frac{R_{f,i}}{T} \left( \beta_i + \frac{T_{a,i}}{T} - \sum_{j=1}^{\numsp} \nu_{ji}^{\prime} \right) \nonumber \\
&& -\: R_{r,i} \left( \frac{1}{T} \left( \beta_i + \frac{T_{a,i}}{T} - \sum_{j=1}^{\numsp} \nu_{ji}^{\prime\prime} \right) - \sum_{j=1}^{\numsp} \nu_{ji} \dydx{B_j}{T} \right) \;. \IEEEeqnarraynumspace
\end{IEEEeqnarray}}%

For all reversible reactions,
{\allowdisplaybreaks \begin{IEEEeqnarray}{rCl}
\dydx{R_i}{Y_j} = \sum_{k=1}^{\numsp} \Biggl( \nu_{ki}^{\prime} \biggl( &&- R_{f,i} W \Bigl( \frac{1}{W_j} - \frac{1}{W_{\numsp}} \Bigr) \nonumber \\
&& +\: \left( \delta_{k j} - \delta_{k \numsp} \right) k_{f,i} \frac{\rho}{W_k} [X_k]^{\nu_{ki}^{\prime} - 1} \prod_{\substack{l=1 \\ l \ne k}}^{\numsp} [X_l]^{\nu_{li}^{\prime}} \biggr) \nonumber \\
&& -\: \nu_{ki}^{\prime\prime} \biggl( -R_{r,i} W \Bigl(\frac{1}{W_j} - \frac{1}{W_{\numsp}} \Bigr) \nonumber \\
&& +\: \left( \delta_{k j} - \delta_{k \numsp} \right) k_{r,i} \frac{\rho}{W_k} [X_k]^{\nu_{ki}^{\prime\prime} - 1} \prod_{\substack{l=1 \\ l \ne k}}^{\numsp} [X_l]^{\nu_{li}^{\prime\prime}} \biggr) \Biggr) \;.
\end{IEEEeqnarray}}%

The partial derivatives of $c_i$ depend on the type of reaction.
For elementary reactions,
\begin{align}
\dydx{c_i}{T} &= 0 \text{ and}\\
\dydx{c_i}{Y_j} &= 0 \;.
\end{align}
For third-body-enhanced reactions,
{\allowdisplaybreaks \begin{IEEEeqnarray}{rCl}
\dydx{c_i}{T} &=&
-\frac{c_i}{T} \text{ , and} \\
\dydx{c_i}{Y_j}
&=& -W c_i \Bigl(\frac{1}{W_j} - \frac{1}{W_{\numsp}} \Bigr) + \rho \Bigl(\frac{\alpha_{ij}}{W_j} - \frac{\alpha_{i \numsp}}{W_{\numsp}} \Bigr) \;.
\end{IEEEeqnarray}}%
Note that in the case that all species contribute equally (i.e., $\alpha_{ij} = 1$ for all species $j$), the latter partial derivative simplifies to
{\allowdisplaybreaks \begin{IEEEeqnarray}{rCl}
\dydx{c_i}{Y_j} = \ddx{Y_j} \left( \frac{p}{\mathcal{R} T} \right) = 0 \;,
\label{e:dcdyequalzero}
\end{IEEEeqnarray}}%
because $\dydx{p}{Y_j} = 0$ (shown in \ref{A:pres_deriv}).

In the case of unimolecular\slash recombination fall-off reactions,
{\allowdisplaybreaks \begin{IEEEeqnarray}{rCl}
\dydx{c_i}{T}
&=& c_i \left( \frac{1}{P_{r,i} (1+P_{r,i})} \dydx{P_{r,i}}{T} + \frac{1}{F_i} \dydx{F_i}{T} \right) \\
\dydx{c_i}{Y_j}
&=& c_i \left( \frac{1}{P_{r,i} (1 + P_{r,i})} \dydx{P_{r,i}}{Y_j} + \frac{1}{F_i} \dydx{F_i}{Y_j} \right) \;. \label{dCidY_falloff}
\end{IEEEeqnarray}}%
The partial derivatives for $P_{r,i}$ are
{\allowdisplaybreaks \begin{IEEEeqnarray}{rCl}
\dydx{P_{r,i}}{T}
&=& \frac{P_{r,i}}{T} \left( \beta_{0,i} - \beta_{\infty,i} + \frac{T_{a 0, i} - T_{a \infty,i}}{T} - 1 \right) \label{E:dPridT} \\
\dydx{P_{r,i}}{Y_j}
 &=& \begin{dcases}
-P_{r,i} W \Bigl(\frac{1}{W_j} - \frac{1}{W_{\numsp}} \Bigr) + \frac{k_{0,i}}{k_{\infty,i}} \rho \Bigl(\frac{\alpha_{ij}}{W_j} - \frac{\alpha_{i \numsp}}{W_{\numsp}} \Bigr) \;, \text{or} \\
-P_{r,i} W \Bigl(\frac{1}{W_j} - \frac{1}{W_{\numsp}} \Bigr) + \frac{k_{0,i}}{k_{\infty,i}} \frac{\rho}{W_m} \left( \delta_{mj} - \delta_{m \numsp} \right) , \; \text{if } [X]_i = [X_m] \;.
\end{dcases}
\label{E:dPridY}
\end{IEEEeqnarray}}%

Similarly, for chemically-activated bimolecular reactions
{\allowdisplaybreaks \begin{IEEEeqnarray}{rCl}
\dydx{c_i}{T} &=& c_i \left( \frac{-1}{1+P_{r,i}} \dydx{P_{r,i}}{T} + \frac{1}{F_i} \dydx{F_i}{T} \right) \\
\dydx{c_i}{Y_j} &=& c_i \left( \frac{-1}{1+P_{r,i}} \dydx{P_{r,i}}{Y_j} + \frac{1}{F_i} \dydx{F_i}{Y_j} \right) \label{dCidY_chem}
\end{IEEEeqnarray}}%

The partial derivatives of $F_i$ depend on the representation of pressure dependence.
For Lindemann reactions,
\begin{align}
  \dydx{F_i}{T} &= 0 \text{ and}\\
  \dydx{F_i}{Y_j} &= 0 \;.
\end{align}
For reactions using the Troe falloff formulation,
\begin{align}
  \dydx{F_i}{T} &= \dydx{F_i}{F_{\text{cent}}} \dydx{F_{\text{cent}}}{T} + \dydx{F_i}{P_{r,i}} \dydx{P_{r,i}}{T} \text{ and} \\
  \dydx{F_i}{Y_j} &=
  \dydx{F_i}{P_{r,i}} \dydx{P_{r,i}}{Y_j} \;,
\end{align}
where
{\allowdisplaybreaks \begin{IEEEeqnarray}{rCl}
  \dydx{F_i}{F_{\text{cent}}} &=& F_i \left( \frac{1}{F_{\text{cent}} \left( 1 + (A_{\text{Troe}} / B_{\text{Troe}})^2 \right)} \right. \nonumber \\
  && \left. -\: \ln F_{\text{cent}} \frac{2 A_{\text{Troe}}}{B_{\text{Troe}}^3} \frac{ \dydx{A_{\text{Troe}}}{F_{\text{cent}}} B_{\text{Troe}} - A_{\text{Troe}} \dydx{B_{\text{Troe}}}{F_{\text{cent}}} }{ \left( 1 + (A_{\text{Troe}} / B_{\text{Troe}})^2 \right)^2} \right), \\
  \dydx{F_{\text{cent}}}{T} &=& -\frac{1-a}{T^{***}} \exp \left( \frac{-T}{T^{***}} \right) - \frac{a}{T^*} \exp \left( \frac{-T}{T^*} \right) + \frac{T^{**}}{T^2} \exp \left( \frac{-T^{**}}{T} \right),  \IEEEeqnarraynumspace \\
  \dydx{F_i}{P_{r,i}} &=& -F_i \ln F_{\text{cent}} \frac{2A_{\text{Troe}}}{B_{\text{Troe}}^3} \frac{ \dydx{A_{\text{Troe}}}{P_{r,i}} B_{\text{Troe}} - A_{\text{Troe}} \dydx{B_{\text{Troe}}}{P_{r,i}} }{ \left( 1 + (A_{\text{Troe}} / B_{\text{Troe}})^2 \right)^2 },
\end{IEEEeqnarray}}%
$\dydx{P_{r,i}}{T}$ is given by Eq.~\eqref{E:dPridT}, $\dydx{P_{r,i}}{Y_j}$ is given by Eq.~\eqref{E:dPridY}, and
{\allowdisplaybreaks \begin{IEEEeqnarray}{rClrCl}
\dydx{A_{\text{Troe}}}{F_{\text{cent}}} &=& \frac{-0.67}{F_{\text{cent}} \log 10} \quad\quad
\dydx{B_{\text{Troe}}}{F_{\text{cent}}} &=& \frac{-1.1762}{F_{\text{cent}} \log 10} \\
\dydx{A_{\text{Troe}}}{P_{r,i}} &=& \frac{1}{P_{r,i} \log 10} \quad\quad
\dydx{B_{\text{Troe}}}{P_{r,i}} &=& \frac{-0.14}{P_{r,i} \log 10} \;.
\end{IEEEeqnarray}}%
Finally, for falloff reactions described with the SRI formulation,
{\allowdisplaybreaks \begin{IEEEeqnarray}{rCl}
  \dydx{F_i}{T} & = & F_i \left( \frac{e}{T} + X \frac{ \frac{a b}{T^2} \exp \left( \frac{-b}{T} \right) - \frac{1}{c} \exp \left( \frac{-T}{c} \right) }{ a \cdot \exp \left( \frac{-b}{T} \right) + \exp \left( \frac{-T}{c} \right) } \right. \nonumber \\
  & & \left. +\: \dydx{X}{P_{r,i}} \dydx{P_{r,i}}{T} \log \left( a \cdot \exp \left( \frac{-b}{T} \right) + \exp \left( \frac{-T}{c} \right) \right) \right) \\
  \dydx{F_i}{Y_j} & = & F_i \dydx{X}{Y_j} \log \left( a \cdot \exp \left( \frac{-b}{T} \right) + \exp \left( \frac{-T}{c} \right) \right)
\end{IEEEeqnarray}}%
where
{\allowdisplaybreaks \begin{IEEEeqnarray}{rCl}
\dydx{X}{P_{r,i}} &=& -X^2 \frac{ 2 \log_{10} P_{r,i} }{ P_{r, i} \log 10 } \;, \\
\dydx{X}{Y_j} &=& \dydx{X}{P_{r,i}} \dydx{P_{r,i}}{Y_j} \;,
\end{IEEEeqnarray}}%
$\dydx{P_{r,i}}{T}$ is given by Eq.~\eqref{E:dPridT}, and $\dydx{P_{r,i}}{Y_j}$ is given by Eq.~\eqref{E:dPridY}.
Note that for both falloff and chemically activated bimolecular reactions, regardless of falloff parameterization, if all species contribute equally to the third-body concentration $[X]_i$ then the partial derivative of $c_i$ with respect to species mass fraction $Y_j$ is zero, since $\dydx{p}{Y_j} = 0$.

The contributions of the logarithmic and Chebyshev pressure-dependent descriptions to the Jacobian matrix entries require separate descriptions since they do not follow the falloff formulation (e.g., falloff factor $F_i$).
Instead, alternative partial derivatives of the reaction rate coefficient with respect to temperature ($\dydx{k_{f,i}}{T}$) must be provided in place of Eq.~\eqref{e:dkfdT} when calculating the partial derivatives of $R_i$ using Eqs.~\eqref{E:dRfdT-orig}, \eqref{e:dRdT_explicit_rev_general}, or \eqref{e:dkrdT_general}.
In both cases the partial derivatives with respect to species mass fractions ($\dydx{k_{f,i}}{Y_j}$) are zero, because the partial derivative of pressure with respect to species mass fraction is zero.
In addition, since this treatment assumes a constant-pressure system, the partial derivative of pressure with respect to temperature is also zero.

For the logarithmic pressure-dependent Arrhenius rate, the partial derivative with respect to temperature is
{\allowdisplaybreaks \begin{IEEEeqnarray}{rCl}
\dydx{k_{f,i}}{T} &=& \left[ \dydx{(\log k_1)}{T} + \left( \dydx{(\log k_2)}{T} - \dydx{(\log k_1)}{T} \right) \frac{\log p - \log p_1}{\log p_2 - \log p_1} \right. \nonumber \\
   & & \left. +\: \left( \log k_2 - \log k_1 \right) \frac{ \dydx{(\log p)}{T} }{\log p_2 - \log p_1} \right] \cdot k_{f,i} \;,
\end{IEEEeqnarray}}%
where $k_1$ and $k_2$ are given by Eqs.~\eqref{e:plog_k1} and \eqref{e:plog_k2}, respectively.
Note that terms such as $k_1$, $k_2$, $p_1$, and $p_2$ are associated with the $i$th reaction alone; we omit the subscript for clarity, and continue this practice with other reaction-specific terms in the following discussion.
The necessary components can be determined as
{\allowdisplaybreaks \begin{IEEEeqnarray}{rCl}
\dydx{(\log k_1)}{T} &=&
\frac{1}{T} \left( \beta_1 + \frac{T_{a,1}}{T} \right) \;, \\
\dydx{(\log k_2)}{T} &=& \frac{1}{T} \left( \beta_2 + \frac{T_{a,2}}{T} \right) \;, \text{and} \\
\dydx{(\log p)}{T} &=& \frac{1}{p} \dydx{p}{T} = 0 \;.
\end{IEEEeqnarray}}%
The final simplified expression can next be constructed, appropriate for use when pressure falls between $p_1$ and $p_2$:
{\allowdisplaybreaks \begin{IEEEeqnarray}{rCl}
\dydx{k_{f,i}}{T} &=& \frac{k_{f,i}}{T} \left[ \beta_1 + \frac{T_{a,1}}{T} + \left( \beta_2 - \beta_1 + \frac{T_{a,2} - T_{a,1}}{T} \right) \frac{ \log p - \log p_1 }{ \log p_2 - \log p_1 } \right] \;.
\end{IEEEeqnarray}}%

Finally, the partial derivatives of the rate coefficient for the $i$th reaction with a Chebyshev rate expression can be evaluated; again, we omit the subscript $i$ for clarity:
{\allowdisplaybreaks \begin{IEEEeqnarray}{rCl}
\dydx{k_f}{T}
  &=& \log(10) \cdot k_f \sum_{i = 1}^{N_T} \sum_{j = 1}^{N_p} \eta_{ij} \, \ddx{T} \left( \mathcal{T}_{i-1} (\tilde{T}) \, \mathcal{T}_{j-1} \left(\tilde{p} \right) \right) \;, \nonumber
\end{IEEEeqnarray}}%
where
{\allowdisplaybreaks \begin{IEEEeqnarray}{rCl}
\ddx{T} \left( \mathcal{T}_{i-1} (\tilde{T}) \, \mathcal{T}_{j-1} \left(\tilde{p} \right) \right) &=& (i-1) U_{i-2} (\tilde{T}) \, \mathcal{T}_{j-1} (\tilde{p}) \, \dydx{\tilde{T}}{T} \nonumber \\
  & & +\: (j-1) \mathcal{T}_{i-1} (\tilde{T}) \, U_{j-2} (\tilde{p}) \, \dydx{\tilde{p}}{T} \;, \\
\dydx{\tilde{T}}{T} &=& \frac{ -2 T^{-2} }{ T_{\max}^{-1} - T_{\min}^{-1} } \;, \\
\dydx{\tilde{p}}{T} &=& \frac{ \frac{2}{p \log(10)} \dydx{p}{T} }{\log_{10} p_{\max} - \log_{10} p_{\min}} = 0 \;,
\end{IEEEeqnarray}}%
and $U_n$ is the Chebyshev polynomial of the second kind of degree $n$, expressed as
\begin{equation}
U_n (x) = \frac{ \sin \left((n+1) \arccos x\right) }{ \sin ( \arccos x ) } \;.
\end{equation}
Thus, the partial derivative of the forward rate coefficient can be expressed as
{\allowdisplaybreaks \begin{IEEEeqnarray}{rCl}
\dydx{k_f}{T} &=& \frac{k_f}{T} \log(10) \sum_{i = 1}^{N_T} \sum_{j = 1}^{N_p} \eta_{ij} \left( (i-1) U_{i-2} (\tilde{T}) \, \mathcal{T}_{j-1} (\tilde{p}) \frac{-2 T^{-1}}{T_{\max}^{-1} - T_{\min}^{-1}} \right) \;.
\end{IEEEeqnarray}}%

The partial derivative of the reverse rate coefficient $k_{r, i}$ with respect to temperature can be found using Eq.~\eqref{e:dkrdT_general} for both pressure-dependent reaction classes.
The partial derivative of $k_{r,i}$ with respect to species mass fractions is zero, for the same reason as the forward rate coefficient.

\subsection{Evaluation of Jacobian matrix}
\label{section:eval}

The Jacobian matrix can be efficiently filled by arranging the evaluation of elements in an appropriate order.
First, evaluate the Jacobian entries for partial derivatives of species equations ($\mathcal{J}_{k+1, 1}$ and $\mathcal{J}_{k+1, j+1}$ for $k, j = 1, \dotsc, \numsp - 1$):
{\allowdisplaybreaks \begin{IEEEeqnarray}{rCl}
\mathcal{J}_{k+1, 1}
&=& \frac{W_k}{\rho} \sum_{i=1}^{N_{\text{reac}}} \nu_{ki} \left[ \dydx{c_i}{T} \left( R_{f,i} - R_{r,i} \right) + c_i \left( \dydx{R_i}{T} + \frac{R_{f,i} - R_{r,i}}{T} \right) \right] \;, \IEEEeqnarraynumspace
\label{e:jac_dydt}
\end{IEEEeqnarray}}%
where all terms were expressed previously; note that $c_i = 1$ and $\dydx{c_i}{T} = 0$ for pressure-dependent reactions expressed via logarithmic interpolations or Chebyshev polynomials.
The remaining columns are given by
{\allowdisplaybreaks \begin{IEEEeqnarray}{rCl}
\mathcal{J}_{k+1, j+1}
&=& \frac{W_k}{\rho} \left( \dot{\omega}_k W\left(\frac{1}{W_j} - \frac{1}{W_{\numsp}} \right) + \mathlarger{\sum_{i=1}^{N_{\text{reac}}}} \nu_{ki} \left( \vphantom{\prod_{\substack{n=1 \\ n \neq l}}^{\numsp}} \dydx{c_i}{Y_j} ( R_{f,i} - R_{r,i} ) \right. \right. \nonumber \\
&& +\: c_i \sum_{l=1}^{\numsp} \Biggl( \nu_{li}^{\prime} \biggl( -R_{f,i} W \Bigl( \frac{1}{W_j} - \frac{1}{W_{\numsp}} \Bigr) \nonumber \\
&& +\: \left( \delta_{l j} - \delta_{l \numsp} \right) k_{f,i} \frac{\rho}{W_l} [X_l]^{\nu_{li}^{\prime} - 1} \prod_{\substack{n=1 \\ n \ne l}}^{\numsp} [X_n]^{\nu_{ni}^{\prime}} \biggr) \nonumber \\
&& -\: \nu_{li}^{\prime\prime} \biggl( -R_{r,i} W \Bigl(\frac{1}{W_j} - \frac{1}{W_{\numsp}} \Bigr) \nonumber \\
&& +\: \left. \left. \left( \delta_{l j} - \delta_{l \numsp} \right) k_{r,i} \frac{\rho}{W_l} [X_l]^{\nu_{li}^{\prime\prime} - 1} \prod_{\substack{n=1 \\ n \ne l}}^{\numsp} [X_n]^{\nu_{ni}^{\prime\prime}} \biggr) \Biggr) \right) \right) \;. \IEEEeqnarraynumspace
%
\label{e:jac_dydy}
\end{IEEEeqnarray}}%

Next, the Jacobian entries for partial derivatives of the energy equation ($\mathcal{J}_{1, 1}$ and $\mathcal{J}_{1, j+1}$ for $j = 1, \dotsc, \numsp - 1$) are given by
{\allowdisplaybreaks \begin{IEEEeqnarray}{rCl}
\mathcal{J}_{1,1}
&=& \frac{-1}{c_p} \left\lbrace \sum_{k=1}^{\numsp - 1} \left[ \left( c_{p,k} - \frac{h_k}{c_p} \dydx{c_p}{T} \right) \frac{W_k \dot{\omega}_k}{\rho} + h_k \mathcal{J}_{k+1,1} \right] \right. \nonumber \\
 && \left. +\: \left( c_{p,\numsp} - \frac{h_{\numsp}}{c_p} \dydx{c_p}{T} \right) \frac{W_{\numsp} \dot{\omega}_{\numsp}}{\rho} + h_{\numsp} \hat{\mathcal{J}}_{\numsp+1,1} \right\rbrace \\
\mathcal{J}_{1,j+1}
  &=& \frac{-1}{c_p} \left( -\frac{c_{p,j} - c_{p, \numsp}}{\rho c_p} \sum_{k=1}^{\numsp} h_k W_k \dot{\omega}_k \right. \nonumber \\
  && \left. +\: \sum_{k=1}^{\numsp - 1} h_k \mathcal{J}_{k+1,j+1} + h_{\numsp} \hat{\mathcal{J}}_{\numsp+1,j+1} \right) ,
\label{e:jac_dt}
\end{IEEEeqnarray}}%
where $\dydx{c_p}{T}$ is given by Eq.~\eqref{E:dcpdT}; $\hat{\mathcal{J}}_{\numsp+1,1}$ and $\hat{\mathcal{J}}_{\numsp+1,j+1}$ are placeholder variables for the last species $\numsp$ calculated in a similar manner as Eqs.~\eqref{e:jac_dydt} and \eqref{e:jac_dydy}.

\subsection{Jacobian matrix sparsity}
\label{S:sparsity}

While some discuss the sparsity of analytically evaluated Jacobian matrices for chemical kinetics~\cite{Lu:2009gh,Perini:2012gy}, for the governing equations of constant-pressure kinetics using mass fractions the Jacobian is actually dense.
This results from the partial derivative of density with respect to species mass fraction, $\partial \rho / \partial Y_j$, that appears in Eq.~\eqref{e:jac_spec}.
As a consequence, the partial derivatives of all species rate-of-change $\dot{Y}_k$ with respect to the $j$th species' mass fraction $Y_j$, or $\mathcal{J}_{k+1, j+1}$, are potentially nonzero.
However, if the constant-volume assumption instead holds true, then $\partial \rho / \partial Y_j = 0$ and Jacobian matrices exhibit the sparsity shown by Perini et al.~\cite{Perini:2012gy}.
In that case, off-diagonal elements of the Jacobian matrix only arise due to direct species interactions and falloff\slash pressure-dependent reactions.
Unfortunately, simulations of low-speed combustion systems typically rely on the constant-pressure assumption~\cite{Law:2006tu}.

Alternatively, the governing equations can be formulated in terms of species molar concentrations rather than mass fractions~\cite{Lu:2009gh,Bisetti:2012jw}.
This eliminates the presence of $\rho$ in the cross-species Jacobian matrix elements and thus significantly increases matrix sparsity for constant-pressure combustion.
We plan to explore this approach for future versions of \texttt{pyJac}.

\section{Optimization of Jacobian evaluation}
\label{S:opt}

Algorithm~\eqref{p:naive_psuedo} presents pseudocode for a simple implementation of the outlined Jacobian evaluation approach based on the equations in Section~\ref{section:eval}.
Although attractive as a straightforward implementation, it neglects potential reuse of computed products and thus unnecessarily recomputes terms.
Furthermore, this formulation produces source code with substantial numbers of lines even for small kinetic models such as GRI-Mech 3.0~\cite{smith_gri-mech_30} (e.g., $\sim$\num{100000} lines).

Without a strategy to enable the reuse of temporary computed products, even reasonably modern compilers (e.g., \texttt{gcc} 4.4.7) struggle to compile the resulting code, resulting in long compilation times, slow execution, and even occasional crashes of the compiler itself.
This section lays out a restructuring of Algorithm~\eqref{p:naive_psuedo}, presented in Algorithm~\eqref{p:updated_pseudo}, that greatly accelerates Jacobian evaluation via reuse of temporary products to reduce computational overhead.
As a side-benefit of this technique, compilation times are greatly reduced and compiler crashes can be avoided.

\begin{algorithm}
\caption{A pseudo-code for simple/naive generation of the chemical Jacobian.}\label{p:naive_psuedo}
\begin{algorithmic}[0]
  \For {$i$ \text{ in } $1$, \dots, $\numsp - 1$}
    \For {$k$ \text{ in } $1$, \dots, $\numreac$}
      \If {$\nu_{i,k} \neq 0$}
        \State Generate code for reaction $k$'s contribution to $\dydx{\dot{Y_i}}{T}$
      \EndIf
    \EndFor
    \For {$j$ \text{ in } $1$, \dots, $\numsp - 1$}
      \For {$k$ \text{ in } $1$, \dots, $\numreac$}
        \If {$\nu_{j,k} \neq 0$}
          \State Generate code for reaction $k$'s contribution to $\dydx{\dot{Y_k}}{Y_j}$
        \EndIf
      \EndFor
    \EndFor
  \EndFor
  \For {$i$ \text{ in } $1$, \dots, $\numsp - 1$}
    \State \text{Generate code for $\dydx{\dot{T}}{Y_i}$}
  \EndFor
  \State \text{Generate code for $\dydx{\dot{T}}{T}$}
\end{algorithmic}
\end{algorithm}

\begin{algorithm}
\caption{A restructured pseudo-code to enable efficient chemical Jacobian evaluation.}\label{p:updated_pseudo}
\begin{algorithmic}[0]
  \For {$k$ \text{ in } $1$, \dots, $\numreac$}
    \State Define reusable products for temperature derivatives\label{op:dt_reusable}
    \For {$i$ in $1$, \dots, $\numsp - 1$}
      \If {$\nu_{i,k} \neq 0$}
        \State Generate code for reaction $k$'s contribution to $\dydx{\dot{Y_i}}{T}$
      \EndIf
    \EndFor
    \State Define reusable products for species derivatives\label{op:dy_reusable}
    \For {$j$ \text{ in } $1$, \dots, $\numsp - 1$}
      \For {$i$ in $1$, \dots, $\numsp - 1$}
        \If {$\nu_{i,k} \neq 0$}
          \State Generate code for $\dydx{\dot{Y_i}}{Y_j}$
        \EndIf
      \EndFor
    \EndFor
  \EndFor
  \For {$i$ \text{ in } $1$, \dots, $\numsp - 1$}
    \State \text{Generate code for $\dydx{\dot{T}}{Y_i}$}
  \EndFor
  \State \text{Generate code for $\dydx{\dot{T}}{T}$}
\end{algorithmic}
\end{algorithm}

\subsection{Accelerating evaluation via reuse of temporary products}

First, examining Eqs.~\eqref{e:jac_dydt}~and~\eqref{e:jac_dydy}, we see that large portions of the Jacobian entries are constant for a single reaction.
For instance, if we define a temporary variable for the $i$th reaction
\begin{equation}
  \Theta_{\partial T,  i} = \frac{1}{\rho}\left[ \dydx{c_i}{T} \left( R_{f,i} - R_{r,i} \right) + c_i \left( \dydx{R_i}{T} + \frac{R_{f,i} - R_{r,i}}{T} \right) \right] \;,
\end{equation}
then Eq.~\eqref{e:jac_dydt} can be rewritten as
\begin{equation}
  \mathcal{J}_{k+1, 1} = W_k \sum_{i=1}^{N_{\text{reac}}} \nu_{ki} \Theta_{\partial T, i} \;.
\end{equation}
Next, instead of summing over all reactions for a single species, we transform Algorithm~\eqref{p:naive_psuedo} to compute the temporary product for a single reaction $i$ and add to all relevant species:
\begin{IEEEeqnarray}{rCl}
  \mathcal{J}_{k+1, 1} \pluseq \nu_{ki} W_k \Theta_{\partial T, i} \quad k = 1, \ldots, \numsp - 1 \;.
\label{e:jac_temperature_update}
\end{IEEEeqnarray}%
and---as introduced in Sec.~\eqref{section:eval}---the placeholder variable $\hat{\mathcal{J}}_{\numsp + 1, 1}$ may be updated analogously to Eq.~\eqref{e:jac_temperature_update}.
In doing so, the temporary product $\Theta_{\partial T,  i}$ must only be evaluated once for each reaction, rather than once for every species in each reaction with a non-zero net production or consumption rate.

Similarly, for Eq.~\eqref{e:jac_dydy} we can define similar temporary products, although more care must be taken to ensure the correctness for all of the different reaction types.
For all reactions $i$, the following temporary product can be defined:
\begin{equation}
  \Theta_{\partial Y, i,\text{ind}} = -\frac{Wc_i}{\rho} \left( R_{f,i}\sum_{l=1}^{\numsp}\nu_{li}^{\prime} - R_{r,i}\sum_{l=1}^{\numsp}\nu_{li}^{\prime\prime}\right) \;.
\end{equation}
The Jacobian entries $\mathcal{J}_{k+1,j+1}$ for a pressure-independent reaction $i$ can then be updated with
{\allowdisplaybreaks \begin{IEEEeqnarray}{rCl}
\mathcal{J}_{k+1,j+1} &\pluseq& \nu_{ki}\frac{W_k}{W_j}
\Biggl[
	\Theta_{\partial Y,i,\text{ind}} \left(1 - \frac{W_j}{W_{\numsp}}\right) +
	c_i \left( \vphantom{\frac{W_j}{W_{\numsp}}} 
		\left( k_{f,i} S_j^{\prime} - k_{r,i} S_j^{\prime\prime} \right) \right. \nonumber \\
&& -\: \left.
  		\frac{W_j}{W_{\numsp}}
  			\left( k_{f,i} S_{\numsp}^{\prime} - k_{r,i} S_{\numsp}^{\prime\prime} \right) \right)
\Biggr] \quad
  k, j = 1, \ldots, \numsp - 1 \;,
\label{e:pindep_jac_species}
\end{IEEEeqnarray}}%
where $S_l^{\prime}$ and $S_l^{\prime\prime}$ are defined (for $l = \{ j, N_{sp} \}$) as
{\allowdisplaybreaks \begin{IEEEeqnarray}{rCl}
S_l^{\prime} &=& \nu_{li}^{\prime} [X_l]^{\nu_{li}^{\prime} - 1} \prod_{\substack{n=1 \\ n \neq l}}^{\numsp}[X_n]^{\nu_{ni}^{\prime}} \text{ and } \\
S_l^{\prime\prime} &=& \nu_{li}^{\prime\prime} [X_l]^{\nu_{li - 1}^{\prime\prime}} \prod_{\substack{n=1 \\ n \neq l}}^{\numsp}[X_n]^{\nu_{ni}^{\prime\prime}} \;.
\end{IEEEeqnarray}}%
and the placeholder variable $\hat{\mathcal{J}}_{\numsp + 1, j+1}$ may be updated as in Eq.~\eqref{e:pindep_jac_species}.
Note that either one or both of $\nu_{ji}^{\prime}$ and $\nu_{ji}^{\prime\prime}$ can be zero, while $\nu_{{\numsp}i}^{\prime}$ and $\nu_{{\numsp}i}^{\prime\prime}$ are typically both zero if the last species is an inert abundant gas, e.g., \ce{N2} for air-fed combustion.
This means that the added $S_l^{\prime}$ and $S_l^{\prime\prime}$ terms are often zero, and may be omitted in such cases.

For third-body enhanced or falloff reactions, the $\dydx{c_i}{Y_j}$ term in Eq.~\eqref{e:jac_dydy} is inherently dependent on the $j$th species; however, certain simplifications can still be made.
First, for any third-body enhanced or falloff reaction where all third-body efficiencies $\alpha_{i,j}$ are unity, the derivative $\dydx{c_i}{Y_j}$ is identically zero as seen in Eq.~\eqref{e:dcdyequalzero}, and the updating scheme defined in Eq.~\eqref{e:pindep_jac_species} can be used.  Otherwise, the appropriate update schemes described below must be employed.

For third-body-enhanced reactions, we define
{\allowdisplaybreaks \begin{IEEEeqnarray}{rCl}
  \Theta_{\partial Y, i, \text{\nth{3}}} &=&
  	\Theta_{\partial Y,i,\text{ind}} + \frac{-Wc_i}{\rho}R_i \nonumber \\
  	&=&
  		-\frac{Wc_i}{\rho}
  		\left(
  			R_{f,i}\left(1 + \sum_{l=1}^{\numsp}\nu_{li}^{\prime} \right) -
  			R_{r,i}\left(1 + \sum_{l=1}^{\numsp}\nu_{li}^{\prime\prime}\right)
  		\right) \;.
\end{IEEEeqnarray}}%
The Jacobian entries $\mathcal{J}_{k+1,j+1}$ for a third-body reaction without falloff dependence can then be updated with
{\allowdisplaybreaks \begin{IEEEeqnarray}{rCl}
\label{e:thd_jac_species}
\mathcal{J}_{k+1,j+1} &\pluseq&
  \nu_{ki}\frac{W_k}{W_j}
  \Biggl[
    \Theta_{\partial Y, i,\text{\nth{3}}}\left(1 - \frac{W_j}{W_{\numsp}}\right) \nonumber \\
&& +\:
    \left(\alpha_{ij} - \alpha_{i \numsp}\frac{W_j}{W_{\numsp}}\right)R_i
    \nonumber \\
&& +\:  c_i \left(
	\left( k_{f,i} S_j^{\prime} - k_{r,i} S_j^{\prime\prime} \right) \vphantom{\frac{W_j}{W_{\numsp}}} \right. \nonumber \\
&& -\: \left. \frac{W_j}{W_{\numsp}}
	\left( k_{f,i} S_{\numsp}^{\prime} - k_{r,i} S_{\numsp}^{\prime\prime} \right)
		  	\right)
  \Biggr] \IEEEeqnarraynumspace
\end{IEEEeqnarray}
for $k, j = 1, \ldots, \numsp - 1$, and the placeholder variable $\hat{\mathcal{J}}_{\numsp + 1, j+1}$ may be updated using Eq.~\eqref{e:thd_jac_species}.

For reactions with falloff dependence, we define a temporary term corresponding to the $P_r$ polynomial in Eqs.~\eqref{dCidY_falloff}~and~\eqref{dCidY_chem}:
\begin{equation}
  \Theta_{P_{r, i}}  =
\begin{dcases}
  \hfil 0 \; & \text{if Lindemann falloff or chemically activated reaction,} \\
  \frac{1.0}{1.0 + P_r} \; & \text{if Troe\slash SRI falloff reaction, or} \\
  \frac{-P_r}{1.0 + P_r} \; & \text{if Troe\slash SRI chemically activated reaction.}
\end{dcases}
\end{equation}
Following this, we define a temporary term for the falloff function derivative $\dydx{F_i}{Y_j}$:
\begin{equation}
\Theta_{F_i} =
\begin{dcases}
  \hfil 0 \; &\text{if Lindemann, } \\
  \hfil \ln F_\text{cent} \frac{2A_{\text{Troe}}}{B_{\text{Troe}}^3}\frac{0.14 A_{\text{Troe}} + B_{\text{Troe}}}{\log 10 (1 + (A_{\text{Troe}} / B_{\text{Troe}})^2)^2} \; &\text{if Troe, or} \\
  \hfil X^2 \frac{2 \log_{10} P_{r,i}}{\log 10} \log \left(a\cdot\exp\frac{-b}{T} + \exp\frac{-T}{c}\right) \; &\text{if SRI.}
\end{dcases}
\end{equation}
Using the above components, we then define the temporary products for falloff reactions:
{\allowdisplaybreaks \begin{IEEEeqnarray}{rCl}
\Theta_{\partial Y, i,\text{falloff}} &=&
	\Theta_{\partial Y, i,\text{ind}} +
	c_i\left(\Theta_{P_{r,i}} + \Theta_{F_i}\right)\left(\frac{-W}{\rho}\right)R_i \nonumber \\
&=&
	\frac{-Wc_i}{\rho}
	\left(
		R_{f,i}\left(\sum_{l}^{\numsp}\nu_{li}^{\prime} \right) -
		R_{r,i}\left(\sum_{l}^{\numsp}\nu_{li}^{\prime\prime}\right) +
		R_i \left(\Theta_{P_{r,i}} + \Theta_{F_{i}} \right)
	\right) \;.
\end{IEEEeqnarray}}%

The Jacobian entries $\mathcal{J}_{k+1, j+1}$ for a falloff dependent reaction $i$ can then be updated as
{\allowdisplaybreaks \begin{IEEEeqnarray}{rCl}
\mathcal{J}_{k+1,j+1} &\pluseq& \nu_{ki}\frac{W_k}{W_j}
\Biggr[
    \Theta_{\partial Y, i, \text{falloff}}\left(1 - \frac{W_j}{W_{\numsp}}\right) \nonumber \\
&& +\:  \frac{\frac{k_0}{k_{\infty}}F_i}{1 + Pr_i}\left(\Theta_{P_{r,i}} + \Theta_{F_i}\right)
		\left(
			\delta_{M}\left(\alpha_{ij} - \alpha_{i \numsp}\frac{W_j}{W_{\numsp}}\right) \right. \nonumber \\
&& +\:
			\left. \left(1 - \delta_{\text{M}}\right)\left(\delta_{mj} - \delta_{m \numsp}\frac{W_j}{W_{\numsp}}\right)
		\right)R_i \nonumber \\
&& +\:  c_i \left(
				\left( k_{f,i} S_j^{\prime} - k_{r,i} S_j^{\prime\prime} \right) -
				\frac{W_j}{W_{\numsp}}
			  			\left( k_{f,i} S_{\numsp}^{\prime} - k_{r,i} S_{\numsp}^{\prime\prime} \right)
		  	\right)
  \Biggr] \IEEEeqnarraynumspace
\label{e:pdep_jac_species}
\end{IEEEeqnarray}}%
for $k, j = 1, \ldots, \numsp - 1$, where $\delta_{\text{M}}$ is defined as:
\begin{equation}
  \delta_M =
\begin{dcases}
1 &\text{if the mixture acts as third body, or} \\
0 &\text{if species $m$ acts as third body.}
\end{dcases}
\end{equation}
The placeholder variable $\hat{\mathcal{J}}_{\numsp + 1, j+1}$ may be updated as in Eq.~\eqref{e:pdep_jac_species}

Finally, all Jacobian entries $\mathcal{J}_{k+1,j+1}$ and placeholder term $\hat{\mathcal{J}}_{\numsp + 1, j+1}$ must be finished with the addition of the species rate term:
{\allowdisplaybreaks \begin{IEEEeqnarray}{rCl}
\mathcal{J}_{k+1,j+1} &\pluseq& \frac{W_k}{W_j} \frac{\dot{\omega}_kW}{\rho}\left(1 - \frac{W_j}{W_{\numsp}}\right) \; \\
\hat{\mathcal{J}}_{\numsp + 1, 1} &\pluseq& \frac{W_{\numsp}}{W_j} \frac{\dot{\omega}_{\numsp}W}{\rho}\left(1 - \frac{W_j}{W_{\numsp}}\right)
\end{IEEEeqnarray}}

As with the update scheme used in Eq.~\eqref{e:jac_temperature_update}, the strategy presented in Eqs.~\eqref{e:pindep_jac_species}, \eqref{e:thd_jac_species}, and~\eqref{e:pdep_jac_species} reduces the computational overhead of Jacobian evaluation.
The bulk of the computation is performed once per reaction, and only minor sub-products must be computed for each species-species pair for a given reaction.

\section{Results and discussion}
\label{S:results}

The Python~\cite{Python:2015} package \texttt{pyJac} implements the methodology for producing analytical Jacobian matrices described in the previous sections, which we released openly online~\cite{Niemeyer:2016py} under the MIT license.
\texttt{pyJac} requires the Python module \texttt{NumPy}~\cite{Walt:2011aa}.
The modules used to test the correctness and performance of \texttt{pyJac} are included in this release, and additionally require \texttt{Cython}~\cite{Behnel:2011aa}, Cantera~\cite{Goodwin:2015aa}, \texttt{PyYAML}~\cite{Simonov:2014aa}, and \texttt{Adept}~\cite{adept-v11}; however, these are not required for Jacobian\slash rate subroutine generation.
In addition, interpreting Cantera-format models~\cite{Goodwin:2015aa} requires installing the Cantera module for any purpose, while \texttt{pyJac} includes native support for interpreting Chemkin-format models~\cite{chemkin:2012}.

In order to demonstrate the correctness and computational performance of the generated analytical Jacobian matrices, we chose four chemical kinetic models as test cases, selected to represent a wide spectrum of sizes, classes of fuel species, and types of reaction rate pressure dependence formulations (e.g., Lindemann\slash Troe\slash SRI falloff formulations, Chebyshev, pressure-log).
Table~\ref{T:models} summarizes the chemical kinetics models used as benchmarks in this work, including the \ce{H2}\slash \ce{CO} model of Burke et al.~\cite{Burke:2011fh}, GRI-Mech 3.0~\cite{smith_gri-mech_30}, USC-Mech version II~\cite{Wang:2007}, and the isopentanol model of Sarathy et al.~\cite{Sarathy:2013jr}.

\begin{table}[tbp]
\centering
\begin{tabular}{@{}l l l c c c c c l@{}}
\toprule
Model & $\numsp$ & $\numreac$ & Lind.\ & Troe & SRI & P-log & Cheb.\ & Ref.\ \\
\midrule
\ce{H2}\slash \ce{CO} & 13 & 27 & & $\times$ & & & & \cite{Burke:2011fh} \\
GRI-Mech.~3.0 & 53 & 325 & $\times$ & $\times$ & & & & \cite{smith_gri-mech_30} \\
USC-Mech II & 111 & 784 & $\times$ & $\times$ & & & & \cite{Wang:2007}\\
\ce{iC5H11OH} & 360 & 2172 & $\times$ & $\times$ & $\times$ & $\times$ & $\times$ & \cite{Sarathy:2013jr} \\
\bottomrule
\end{tabular}
\caption{
Summary of chemical kinetic models used as benchmark test cases. All models contain third-body reactions with enhanced third bodies. The ``Lind.'', ``Troe'', ``SRI'', ``P-log'', and ``Cheb.'' columns indicate the presence of Lindemann, Troe, and SRI falloff; logarithmic pressure interpolation; and Chebyshev pressure-dependent reactions, respectively. Two reactions with photons were removed from the Sarathy et al.~\cite{Sarathy:2013jr} model since neither Cantera nor \texttt{pyJac} supports such reactions.
}
\label{T:models}
\end{table}

For both the correctness and performance tests, stochastic partially stirred reactor (PaSR) simulations generated thermochemical composition data covering a wide range of temperatures and species mass fractions.
\ref{A:pasr} contains a detailed description of the PaSR methodology and implementation; in addition, the \texttt{pyJac} package contains the PaSR code used in the present study.
We performed nine premixed PaSR simulations for each fuel, with the parameters given in Table~\ref{T:pasr_parameters}.
Each simulation ran for five (isopentanol) or ten (hydrogen, methane, and ethylene) residence times ($\tau_{\text{res}}$) to ensure reaching statistical steady state.

\begin{table}[tbp]
\centering
\begin{tabular}{@{}l l l l l@{}}
\toprule
Parameter & \ce{H2} & \ce{CH4} & \ce{C2H4} & \ce{iC5H11OH} \\
\midrule
$\phi$ & \multicolumn{4}{c}{1} \\
$T_{\text{in}}$ & \multicolumn{4}{c}{\SIlist{400;600;800}{\kelvin}} \\
$p$ & \multicolumn{4}{c}{\SIlist{1;10;25}{\atm}} \\
$N_{\text{part}}$ & \multicolumn{4}{c}{100} \\
$\tau_{\text{res}}$ & \SI{10}{\milli\second} & \SI{5}{\milli\second} & \SI{100}{\micro\second} & \SI{100}{\micro\second} \\
$\tau_{\text{mix}}$ & \SI{1}{\milli\second} & \SI{1}{\milli\second} & \SI{10}{\micro\second} & \SI{10}{\micro\second} \\
$\tau_{\text{pair}}$ & \SI{1}{\milli\second} & \SI{1}{\milli\second} & \SI{10}{\micro\second} & \SI{10}{\micro\second} \\
$N_{\text{res}}$ & 10 & 10 & 10 & 5 \\
\bottomrule
\end{tabular}
\caption{
PaSR parameters used for hydrogen\slash air, methane\slash air, ethylene\slash air, and isopentanol\slash air premixed combustion cases, where $\phi$ indicates equivalence ratio, $T_{\text{in}}$ is the temperature of the inflowing particles, $p$ is the pressure, $N_{\text{part}}$ is the number of particles, $\tau_{\text{res}}$ is the residence time, $\tau_{\text{mix}}$ is the mixing time, $\tau_{\text{pair}}$ is the pairing time, and $N_{\text{res}}$ is the number of residence times simulated.
}
\label{T:pasr_parameters}
\end{table}

\subsection{Validation}
\label{s:validation}

In order to test the correctness of the Jacobian matrices produced by \texttt{pyJac}, we initially compared the resulting analytical matrices against numerical approximations based on finite differences.
However, while these commonly provide numerical approximations to Jacobian matrices, potential scaling issues due to large disparities in species mass fractions and associated net production rates can cause challenges in selecting appropriate differencing step sizes~\cite{Shampine:1994aa}.
For example, we encountered large errors in some partial derivatives under certain conditions when initially attempting to evaluate derivatives using sixth-order central differences with increments calculated similarly to the implicit CVODE integrator~\cite{Hindmarsh:2005hg}.
Adjusting the relative and absolute tolerances reduced some of these errors, but we could not find consistent, effective values for all the states considered---in particular, states near equilibrium required unreasonably small tolerances (e.g., \num{1e-20}), but these same tolerances caused larger errors at other states.
Note that the discussion of our experiences are not intended to imply issues with the numerical Jacobian approach used in CVODE or other implicit integrators; such solvers only require approximations to the Jacobian matrix.
However, based on our experiences, we recommend taking care when attempting to obtain highly accurate Jacobian matrices, as in the current effort and for analysis techniques such as computational singular perturbation~\cite{Lam:1988wc,Lam:1993ub,Lam:1994ws,Lam:2007jq} and chemical explosive mode analysis~\cite{LU:2010hp,Luo:2012hk,Shan:2012dy} that rely on eigendecomposition of the Jacobian matrix.

As a result of the aforementioned difficulties with finite differences, we obtained accurate Jacobian matrices via automatic differentiation through expression templates using the \texttt{Adept} software library~\cite{hogan2014fast,adept-v11}.
In our experience, numerical Jacobians obtained using multiple-term Richardson extrapolants~\cite{Richardson:1911cl,Lyness:1966aa,Lyness:1969bt} are of similar accuracy but have a much higher computational cost due to the large number of function evaluations required.
We did not explore other options for obtaining highly accurate numerical Jacobian approximations using, e.g., complex-step derivatives~\cite{Martins:2003fv,Shampine:2007gw,Ridout:2009fn}.
For all test cases, the correctness of the species and reaction rate subroutines were established through comparison with Cantera~\cite{Goodwin:2015aa}.

In reporting the discrepancy between the analytical and automatically differentiated Jacobian matrices, denoted by $\mathcal{J}$ and $\hat{\mathcal{J}}$ respectively, we used the Frobenius norm of the relative errors of matrix elements
\begin{equation}
E_\text{rel} = \norm{ \frac{ \hat{\mathcal{J}} - \mathcal{J} }{ \hat{\mathcal{J}} } }_F \;,
\label{e:error_norm}
\end{equation}
to quantify error.
This differs somewhat from the error metric suggested by Anderson et al.~\cite{Anderson:1999aa} for use quantifying the error of matrices in LAPACK: the relative error Frobenius norm
\begin{equation}
E_\text{norm} = \frac{ \norm{ \hat{\mathcal{J}} - \mathcal{J} }_F }{ \norm{\hat{\mathcal{J}}}_F } \;.
\end{equation}
$E_\text{norm}$ indicates overall error in the matrix but can be dominated by large elements, so we believe the error measure $E_\text{rel}$ given by Eq.~\eqref{e:error_norm} to be more useful in identifying discrepancies in both larger and smaller matrix elements.

Table~\ref{T:error} presents the correctness testing results for each benchmark case.
For certain conditions, we observed large relative errors in small nonzero elements (e.g., magnitudes of $\sim$\num{e-22}) likely due to roundoff errors; thus, error statistics comprised only matrix elements $\mathcal{J}_{ij}$ where $\abs{\mathcal{J}_{ij}} \geq \norm{\mathcal{J}}_F / \num{e20}$.
For all benchmark cases, the analytical Jacobian matrices match closely with those obtained via automatic differentiation, with the largest discrepancy below \SI{1e-3}{\percent}.

\begin{table}[tbp]
\centering
\begin{tabular}{@{}l l l l@{}}
\toprule
Model                 & Sample size   & Mean (\percent) & Maximum (\percent) \\
\midrule
\ce{H2}\slash \ce{CO} & \num{900900}  & \num{3.170e-8}  & \num{9.828e-4} \\
GRI-Mech 3.0          & \num{450900}  & \num{2.946e-8}  & \num{1.240e-4} \\
USC-Mech II           & \num{91800}	  & \num{9.046e-9}  & \num{2.356e-4} \\
\ce{iC5H11OH}         & \num{450900}  & \num{8.686e-10} & \num{1.680e-5} \\
\bottomrule
\end{tabular}
\caption{Summary of Jacobian matrix correctness study results.
Error statistics are based on the norm of the relative error percentages $E_\text{rel}$ for each sample.
The sample size for each case depends on the number of particles and relevant timescales used.
}
\label{T:error}
\end{table}

We also initially made comparisons with the \texttt{TChem} package~\cite{Safta:2011vn,TChem:v0.2}
for validation purposes, using the \ce{H2}\slash \ce{CO} model.
However, while many quantities calculated via \texttt{TChem} agreed closely with those from both
Cantera and \texttt{pyJac}, we observed discrepancies in the \texttt{TChem} species net
production rates that correlated with significant errors in the derivative source term and
Jacobian matrix. These discrepancies occurred mainly for low species production rates, where the
(larger) rates for other species agreed closely, and in general for near-equilibrium states.
Furthermore, during our testing it became apparent that \texttt{TChem} v0.2 is not thread-safe
when parallelized with OpenMP. We discuss this issue further in \ref{A:tchem}.
As a result, we do not present detailed Jacobian matrix comparisons between \texttt{TChem}
and \texttt{pyJac} here.

\subsection{Performance analysis}
\label{s:performance}

The performance of \texttt{pyJac}-generated subroutines for CPU and GPU execution was tested by evaluating Jacobian matrices for the four kinetic models using the previously discussed PaSR thermochemical composition data.
In both cases, the performance of \texttt{pyJac} was compared with that of a finite-difference-based Jacobian; in addition, the CPU routines were compared with \texttt{TChem}~\cite{TChem:v0.2} for the \ce{H2}\slash \ce{CO}, GRI-Mech 3.0, and USC-Mech II models.
The CPU Jacobian subroutines were compiled using \texttt{gcc} 4.8.5~\cite{gcc-4.8.5}, and run on four ten-core Intel Xeon E5-4640 v2 processors with \SI{20}{\mega\byte} of L3 cache memory, installed on an Ace Powerworks PW8027R-TRF+ with a Supermicro X9QR7-TF+/X9QRi-F+ baseboard.
As previously mentioned, our tests found that \texttt{TChem} v0.2 is not thread-safe (c.f., \ref{A:tchem}).
Therefore, all CPU performance comparisons between \texttt{pyJac}, \texttt{TChem}, and the
finite-difference method were carried in a single-threaded manner.
Reported mean evaluation times were computed from the wall-clock run times of ten individual calculations.
The GPU Jacobian subroutines were compiled using \texttt{nvcc} 7.5.17~\cite{cuda-7.5} and tested on a single NVIDIA Tesla C2075 GPU.
The mean GPU evaluation times were again computed from ten individual runs.
For the CPU and GPU finite-difference-based Jacobian calculations, we used a simple first-order forward difference adapted from CVODE~\cite{Hindmarsh:2005hg} to give a realistic comparison with a commonly used finite-difference technique.

Figure~\ref{F:cpu_perf} shows the performance of the CPU-based \texttt{pyJac} Jacobian matrix evaluations for the four kinetic models, compared with the performance of finite difference calculations and the \texttt{TChem} package~\cite{Safta:2011vn}.
(\texttt{TChem} does not support pressure-log or Chebyshev pressure-dependent reaction expressions, so we could not evaluate its performance with the isopentanol model.)
Table~\ref{t:cpu_comp} summarizes the performance ratio between Jacobian matrices calculated using \texttt{pyJac} and finite difference\slash \texttt{TChem}; \texttt{pyJac}-based routines perform approximately \SIrange{1.1}{2.2}{$\times$} faster than \texttt{TChem} on a single-threaded basis, depending on the size of the model.
However, we again note that \texttt{TChem}'s lack of thread-safety implies that calculation of Jacobians for many different thermo-chemical states on a multicore CPU may be much faster using \texttt{pyJac} as the calculations may be accelerated easily using OpenMP.
\texttt{pyJac} also performs \SIrange{3}{7.5}{$\times$} faster than the first-order finite difference technique.

\begin{figure}[tbp]
    \centering
    \includegraphics[width=0.75\textwidth]{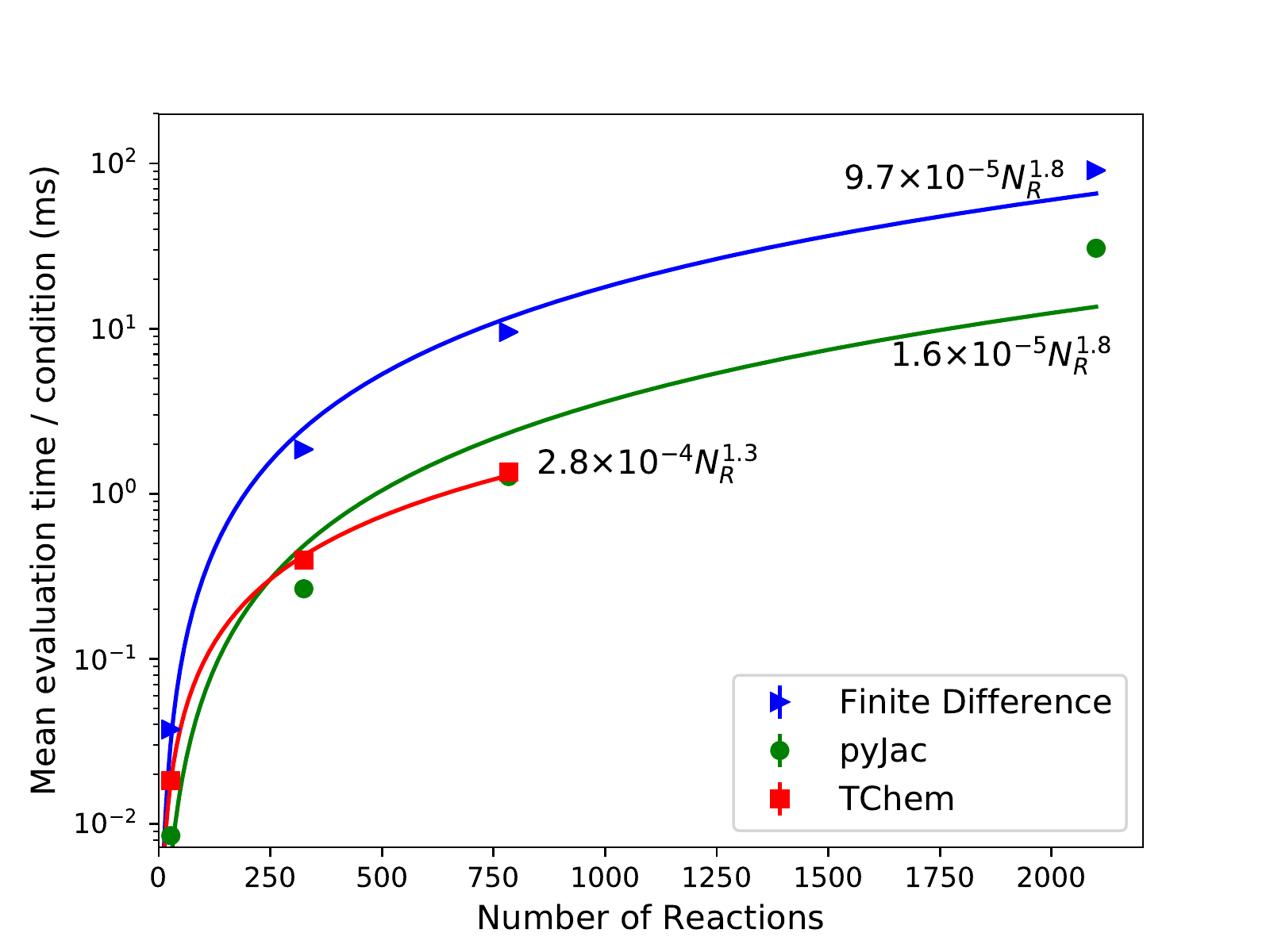}
    \caption{CPU-based Jacobian matrix evaluation times using finite differences, \texttt{pyJac}, and \texttt{TChem} for the four kinetic models.
    No performance data were available for the isopentanol model using \texttt{TChem} due to the presence of unsupported reaction types.
    The symbols indicate performance data, while the solid lines represent least-squares best fits based on the number of reactions $N_R$ in the models.
    Error bars are present, but too small to detect.
    Data, plotting scripts, and the figure file are available under CC-BY~\cite{paperscript:2017}.
    }
    \label{F:cpu_perf}
\end{figure}

\begin{table}[tbp]
\centering
\begin{tabular}{@{}l l l@{}}
\toprule
Model & $\bar{R}_{\texttt{TChem}} / \bar{R}_{\texttt{pyJac}}$ & $\bar{R}_{\text{FD}} / \bar{R}_{\texttt{pyJac}}$ \\
\midrule
\ce{H2}\slash \ce{CO} & 2.16 & 4.40 \\
GRI-Mech 3.0 &  1.49 & 6.98 \\
USC-Mech II &  1.07 & 7.51 \\
\ce{iC5H11OH} & $-$ & 2.96 \\
\bottomrule
\end{tabular}
\caption{Ratio between CPU-based Jacobian matrix evaluation times of \texttt{TChem} and finite difference (FD), and \texttt{pyJac}.
$\bar{R}$ indicates the mean evaluation time.}
\label{t:cpu_comp}
\end{table}

Figure~\ref{F:cpu_perf} also shows best-fit lines based on least-squares regressions for the available data (the corresponding $R^2$ values were 0.89, 0.57, and 0.99 for the finite difference, \texttt{pyJac} and \texttt{TChem} results, respectively).
These fits suggest that \texttt{pyJac} and the finite difference method scale superlinearly---nearly quadratically---with number of reactions, while \texttt{TChem} scales nearly linearly with the number of reactions.
This is expected for the finite difference approach, but warrants explanation for the analytical Jacobian methods.
Lu and Law argued~\cite{Lu:2009gh} that the cost of analytical Jacobian evaluation should scale linearly with the number of reactions in a model.
However, this argument is predicated on the assumption of a sparse Jacobian resulting from a molar-concentration-based system; namely, that only changes in species concentrations that participate in a reaction affect the resulting reaction rate.
As discussed in Section~\ref{S:sparsity}, the constant-pressure\slash mass-fraction-based system used in \texttt{pyJac}---along with the majority of reactive-flow modeling descriptions---results in a dense Jacobian and thus execution times that exhibit a higher-order dependence on the number of species\slash reactions.
The current performance, while faster than either a typical finite difference approach or \texttt{TChem}, motivates developing a sparse-concentration-based constant-pressure Jacobian system in future versions of \texttt{pyJac}.

\begin{figure}[tbp]
    \centering
    \begin{subfigure}{0.75\textwidth}
        \centering
        \includegraphics[width=\textwidth]{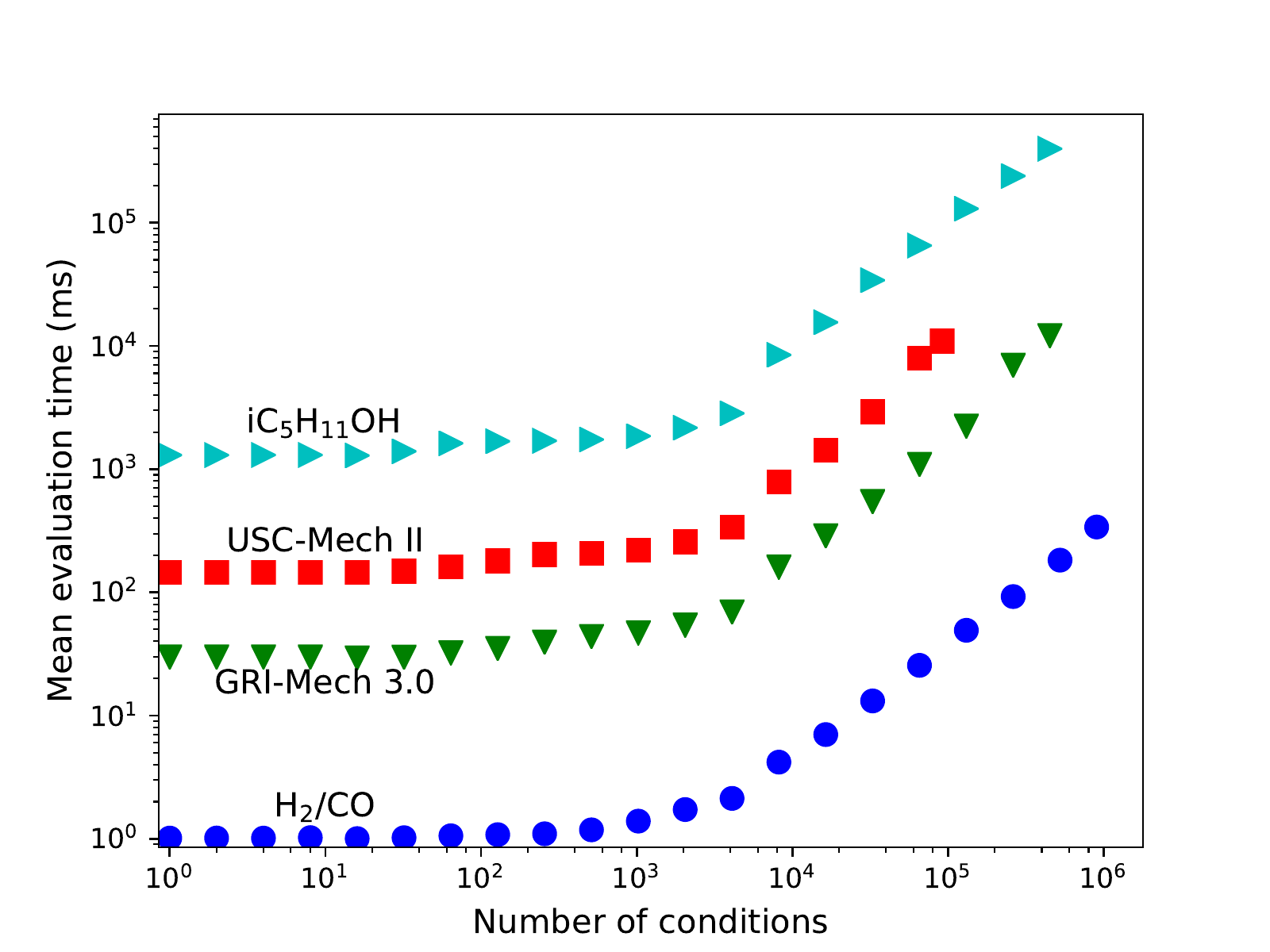}
        \caption{Mean GPU runtime versus the number of conditions evaluated.}
        \label{F:gpu_mean}
    \end{subfigure}%
    \\
    \begin{subfigure}{0.75\textwidth}
        \centering
        \includegraphics[width=\textwidth]{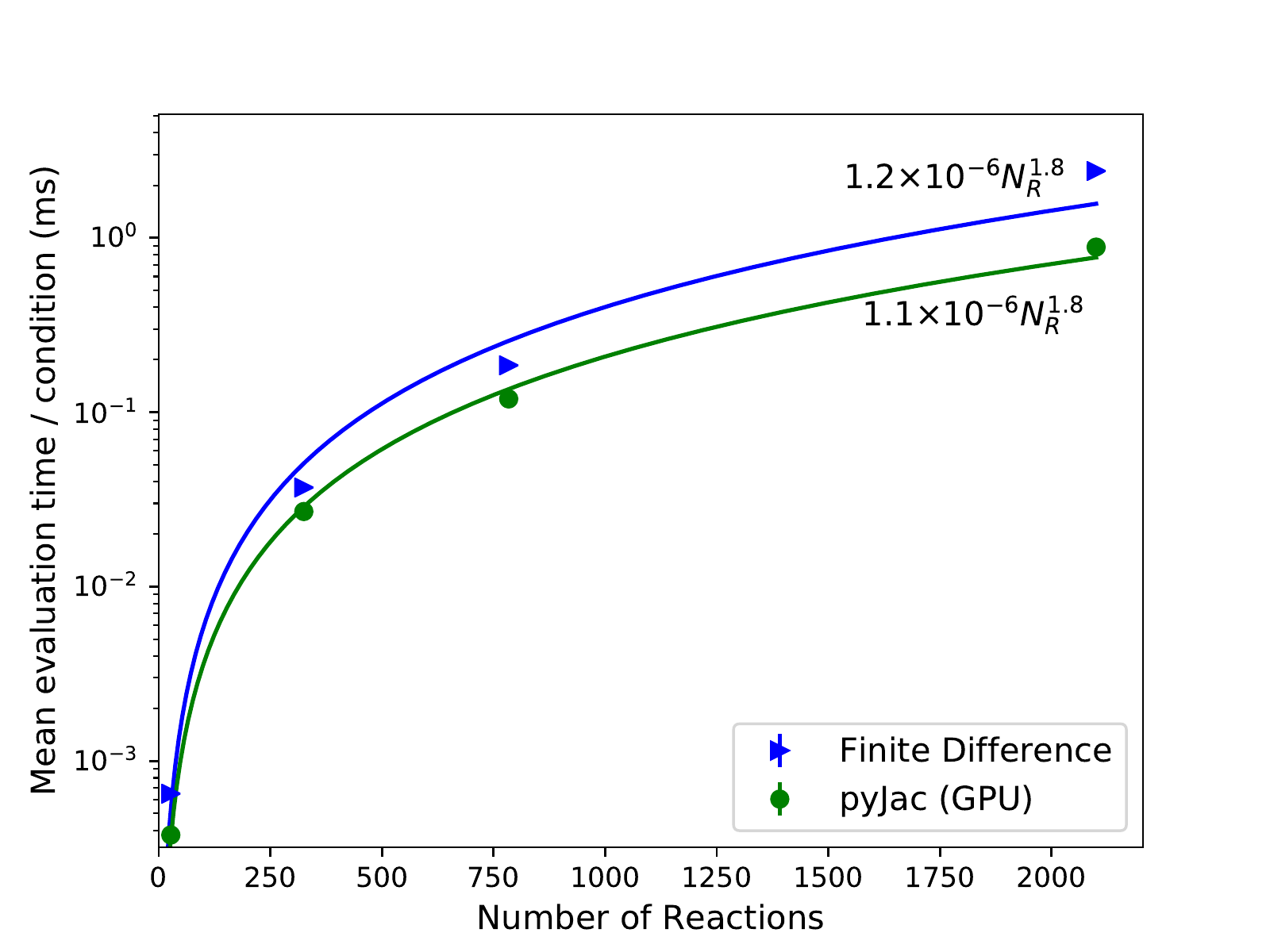}
        \caption{Comparison of normalized GPU \texttt{pyJac} and finite difference Jacobian matrix evaluation times versus the number of reactions in the model. Symbols indicate performance data, while solid lines represent least-squares best fits based on the number of reactions $N_R$ in the models.}
        \label{F:gpu_norm}
    \end{subfigure}
    \caption{Performance of the GPU-based \texttt{pyJac}.
    Note the logarithmic scales of the ordinate axes.
    Error bars are present, but too small to detect.
    Data, plotting scripts, and figure files are available under CC-BY~\cite{paperscript:2017}.
    }
    \label{F:gpu_perf}
\end{figure}

Figure~\ref{F:gpu_perf} demonstrates the performance of the GPU Jacobian matrix implementations.
Figure~\ref{F:gpu_mean} shows the mean runtime of the GPU Jacobian matrix evaluations against the number of conditions---i.e., the number of thermochemical composition states, with one Jacobian matrix evaluated per state.
As the number of conditions increases, the GPU becomes fully utilized and the growth rate of the evaluation time begins increasing linearly (here displayed on a log-log plot).
This thread saturation point occurs at nearly the same number of conditions for each model.
We adopted a ``per-thread'' GPU parallelization model in this work, where a single GPU thread evaluates a single Jacobian matrix, rather than a ``per-block'' model where GPU threads in a block cooperate to evaluate a Jacobian.
The per-thread approach simplifies generation of highly optimized code for SIMD processors and provides a higher theoretical bound on the number of Jacobian matrices that can be evaluated in parallel.
However, the choice of GPU parallelization merits further investigation because memory bandwidth limits the current per-thread implementation due to small cache sizes available on GPU streaming multiprocessors.
A per-block implementation might utilize this small cache more effectively, but it is unclear if this would increase overall performance.
Figure~\ref{F:gpu_norm} shows the longest \texttt{pyJac} and forward finite-difference evaluation times---normalized by the number of conditions---for each kinetic model plotted against the number of reactions in the model.
As with the CPU matrix evaluations shown in Fig.~\ref{F:cpu_perf}, we observed a superlinear (but subquadratic) scaling of the performance with number of reactions.
The least-squares best-fit lines for the \texttt{pyJac} and finite-difference results---with $R^2$ values of 0.98 and 0.82, respectively---exhibit similar polynomial orders to the CPU-based Jacobian evaluation, suggesting dependence on the methods rather than hardware.

Table~\ref{t:gpu_comp} shows the ratios of average evaluation times between finite difference and \texttt{pyJac} on the GPU for the longest runtimes (i.e., the full number of conditions for each model).
Interestingly, \texttt{pyJac} performs noticeably better than the finite-difference technique with the isopentanol model ($\sim$\SI{3}{$\times$}) than the three smaller models (\SIrange{1}{2}{$\times$}).
This likely occurs due to the small cache size of the GPU; in this case, the larger model size (360 species compared to, e.g., 111 species for USC-Mech II) makes it difficult to store a majority of species concentrations in the cache, forcing more loads from global memory.
Thus, using an analytical Jacobian formulation increases performance for larger models by a factor of \numrange{1.6}{2.0} over finite difference methods.
A sparse Jacobian formulation would similarly greatly benefit GPU evaluation due to the greatly reduced memory traffic requirements (i.e., reduced number of global reads and writes).

\begin{table}[tbp]
\centering
\begin{tabular}{@{}l l@{}}
\toprule
Model & $\bar{R}_{\text{FD}} / \bar{R}_{\texttt{pyJac}}$ \\
\midrule
\ce{H2}\slash \ce{CO} & 1.72 \\
GRI-Mech 3.0 &  1.37 \\
USC-Mech II &  1.56 \\
\ce{iC5H11OH} & 2.73 \\
\bottomrule
\end{tabular}
\caption{Ratio between finite difference (FD) and \texttt{pyJac} Jacobian matrix evaluation times on the GPU.
$\bar{R}$ indicates the mean evaluation time.}
\label{t:gpu_comp}
\end{table}

\begin{figure}[tbp]
    \centering
    \begin{subfigure}{0.75\textwidth}
        \centering
        \includegraphics[width=\textwidth]{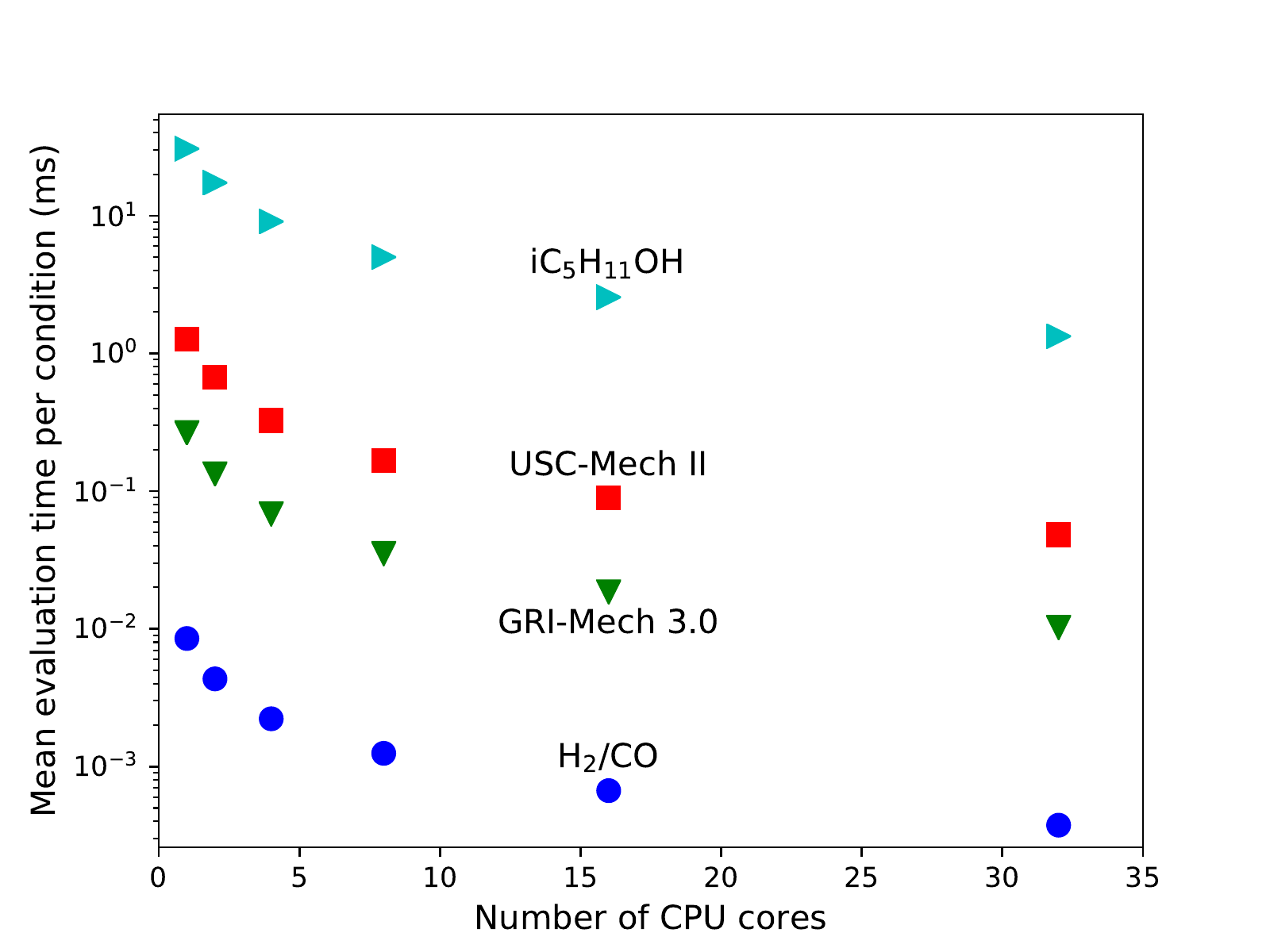}
        \caption{Mean CPU runtime versus the number of CPU cores.
        Error bars are present, but too small to detect. Note the logarithmic
        scales of the ordinate axis.}
        \label{F:cpu_scaling_perf}
    \end{subfigure}%
    \\
    \begin{subfigure}{0.75\textwidth}
        \centering
        \includegraphics[width=\textwidth]{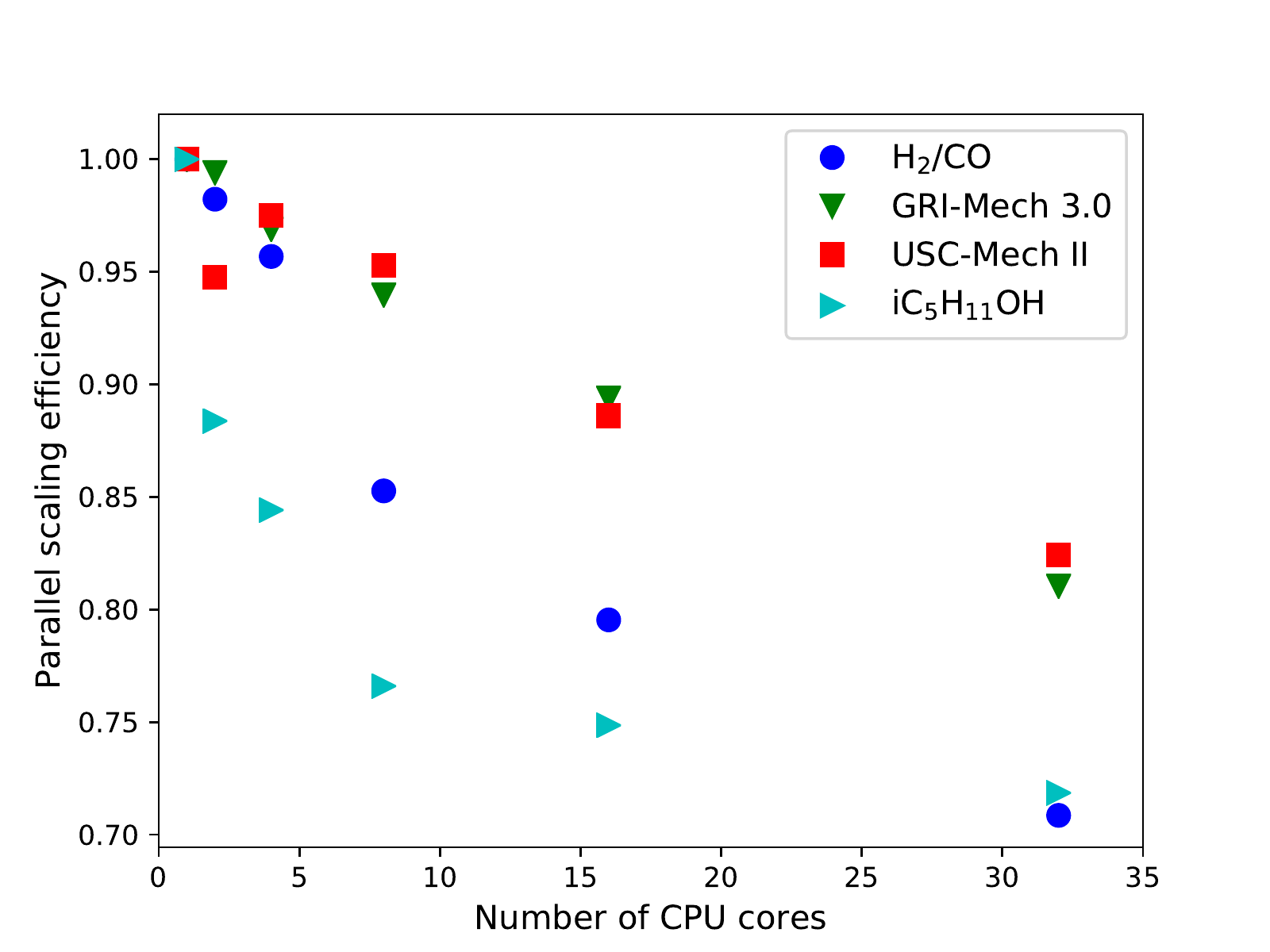}
        \caption{Strong scaling efficiency versus number of CPU cores.}
        \label{F:cpu_scaling_perf_par}
    \end{subfigure}
    \caption{Parallel performance scaling of the CPU-based \texttt{pyJac},
    with the number of CPU cores varying and the number of states fixed
    at the value associated with each model given in Table~\ref{T:error}.
    Data, plotting scripts, and figure files are available under CC-BY~\cite{paperscript:2017}.
    }
    \label{F:cpu_scaling}
\end{figure}

Lastly, Fig.~\ref{F:cpu_scaling} shows the scaling of performance with the
number of CPU cores, ranging from 1 to 32, and the number of conditions fixed at
the value given for each model in Table~\ref{T:error}.
Figure~\ref{F:cpu_scaling_perf} shows the mean evaluation time, which exhibits
a power-law dependence on the number of cores. The functional relationship
between evaluation time and number of cores is similar for each model, with an
intercept increasing with model size. Figure~\ref{F:cpu_scaling_perf_par}
shows the strong scaling efficiency with number of cores. Interestingly,
the strong scaling efficiency drops most significantly for the smallest and
largest models (\ce{H2}\slash \ce{CO} and \ce{iC5H11OH}, respectively) with
increasing number of processors. We hypothesize that the reasons for the sharp
drop in scaling efficiency of these two models actually differ. The cost of
evaluation \ce{H2}\slash \ce{CO} is low enough, particularly at increasing numbers of CPU
cores, that the relative overhead for launching OpenMP threads is larger and
thus the efficiency drops. Conversely, the much larger size of the \ce{iC5H11OH}
model requires frequent accesses of larger amounts of memory, and thus possibly
also more frequent cache misses.

The results presented above raise a number of issues that warrant further study.
The superlinear scaling of \texttt{pyJac} matrix evaluation time with number of reactions in the kinetic models results from the density of the Jacobian matrix, and suggests the potential benefits of a sparse molar-concentration-based Jacobian formulation.
However, while \texttt{pyJac} outperforms \texttt{TChem} in all cases, the performance ratio between them reduces with increasing number of reactions, suggesting that the performance benefit of a hard-coded subroutine may decrease with increasing model size.
Although it appears from these results that the performance of a hard-coded Jacobian subroutine can surpass that of an interface-based Jacobian evaluation, the hard-coded method has additional difficulties due to the size and number of files created.
For example, the CPU Jacobian evaluation subroutine for USC-Mech II comprises over \num{360000} total lines spread across 22 files (not including the supporting files for, e.g., species and reaction rate evaluations)---attempting to incorporate the entire matrix evaluation in a single file crashed an older version of the \texttt{gcc} compiler (4.4.7).
This not only causes longer source code compilation times (on the order of ten minutes to an hour when using 24 parallel compiler instances), but could cause compiler crashes and inexplicable behavior, particularly for the less mature \texttt{nvcc} compiler.
While we expect a sparse Jacobian formulation would significantly alleviate these issues, future work should directly compare the performance of comparable hard-coded and interface-based Jacobian evaluation approaches.

\section{Conclusions}
\label{S:conclusions}

This work developed the theory behind an analytical Jacobian matrix evaluation
approach for constant-pressure chemical kinetics, including new derivations of
partial derivatives with respect to modern pressure-dependent reaction formulations.
In addition, we detailed strategies for efficient evaluation of the Jacobian
matrix, including reordering matrix element evaluations in order to enable the
reuse of temporary products.
We implemented the presented methodology in the open-source software package
\texttt{pyJac}~\cite{Niemeyer:2016py}, which generates custom source code files
for evaluating chemical kinetics Jacobian matrices on both CPU and GPU systems.
We established the correctness of the resulting Jacobian matrices through
comparison with matrices obtained by automatic differentiation.
The results agree within \SI{0.001}{\percent} for kinetic models describing the
oxidation of hydrogen, methane, ethylene, and isopentanol, with 13 to 360
species (and 27 to 2172 reactions).
Finally, we investigated the performance of the CPU and GPU matrix evaluation
subroutines by evaluating the matrix calculation time for the same kinetic models.
The \texttt{pyJac} CPU-based Jacobian evaluation performs \numrange{3}{7.5}
times faster than a forward-finite-difference approach, while a more modest
improvement speedup of \SIrange{1.1}{2.2}{$\times$} was found compared to
the existing \texttt{TChem} software on a single-threaded basis.
We note again that \texttt{TChem} is \textbf{not} thread-safe when parallelized
with OpenMP,
while \texttt{pyJac} is easily parallelized with the same approach;
for multicore CPUs \texttt{pyJac} is expected to greatly outperform \texttt{TChem}.

Planned development work for \texttt{pyJac} includes adding support for the constant-volume assumption, and the capability to generate Fortran and Matlab source codes.
Further study will be performed into the benefits of the current custom source-code, compiled Jacobian evaluation subroutine approach versus a loop\slash branch-based approach, in order to determine how to obtain the best performance scaling with kinetic model size.
Finally, we will investigate the performance benefits of converting from a differential system based on species mass fractions to molar concentrations, which offers a marked improvement in matrix sparsity.

\section*{Acknowledgments}

This material is based upon work supported by the National Science Foundation under grants ACI-1535065 and ACI-1534688.
In addition, we thank James Sutherland and Mike Hansen of the University of Utah for helpful discussions on appropriate formulation of state vectors.


\appendix
\setcounter{figure}{0}

\renewcommand*{\thesection}{\appendixname~\Alph{section}}

\section{Supplementary material}

The results for this paper were obtained using \texttt{pyJac} v1.0.2~\cite{Niemeyer:2016py}.
The most recent version of \texttt{pyJac} can be found at its GitHub
repository \url{https://github.com/SLACKHA/pyJac}. In addition, the repository
contains detailed documentation and an issue-tracking system.
All figures, and the data and plotting scripts necessary to reproduce them, are
available openly under the CC-BY license~\cite{paperscript:2017}.

\section{Proof of partial derivative of pressure}
\label{A:pres_deriv}

{\allowdisplaybreaks \begin{IEEEeqnarray}{rCl}
\dydx{p}{Y_j} &=& \ddx{Y_j} \left( \mathcal{R} T \sum_{k=1}^{\numsp} [X_k] \right) = \mathcal{R} T \sum_{k=1}^{\numsp} \dydx{[X_k]}{Y_j} \nonumber \\
  &=& \mathcal{R} T \sum_{k=1}^{\numsp} \left[ -[X_k] W \left(\frac{1}{W_j} - \frac{1}{W_{\numsp}} \right) + \left( \delta_{kj} - \delta_{k \numsp} \right) \frac{\rho}{W_k} \right] \nonumber \\
  &=& -\mathcal{R} T W \left( \frac{1}{W_j} - \frac{1}{W_{\numsp}} \right) \sum_{k=1}^{\numsp} [X_k] + \mathcal{R} T \rho \sum_{k=1}^{\numsp} \frac{\delta_{kj} - \delta_{k \numsp}}{W_k} \nonumber \\
  &=& -pW \left( \frac{1}{W_j} - \frac{1}{W_{\numsp}} \right) + \rho \mathcal{R} T \left( \frac{1}{W_j} - \frac{1}{W_{\numsp}} \right) \nonumber \\
  &=& \left(-p W + pW \right) \left( \frac{1}{W_j} - \frac{1}{W_{\numsp}} \right) \nonumber \\
\therefore \dydx{p}{Y_j} &=& 0 \;.
\end{IEEEeqnarray}}%
Note that this holds without any assumption of constant pressure or volume.

\section{Partially stirred reactor implementation}
\label{A:pasr}

Here, we describe our partially stirred reactor (PaSR) implementation for completeness, based on prior descriptions~\cite{Correa:1993ud,Chen:1997ta,Pope:1997wu,Bhave:2004hc,Ren:2004fz,Ren:2014cd}.
The reactor model consists of an even number $N_{\text{part}}$ of particles, each with a time-varying composition $\phi (t)$.
Unlike the composition vector given previously (Eq.~\eqref{e:vars}), here we use mixture enthalpy and species mass fractions to describe the state of a particle:
\begin{equation}
\phi = \left \lbrace h, Y_1, Y_2, \dotsc, Y_{\numsp} \right \rbrace^{\intercal} \;.
\end{equation}
At discrete time steps of size $\Delta t$, events including inflow, outflow, and pairing cause certain particles to change composition; between these time steps, mixing and reaction fractional steps separated by step size $\Delta t_{\text{sub}}$ evolve the composition of all particles.

Inflow and outflow events at the discrete time steps comprise the inflow stream compositions replacing the compositions of $N_{\text{part}} \Delta t / \tau_{\text{res}}$ randomly selected particles, where $\tau_{\text{res}}$ is the residence time.
For premixed combustion cases, the inflow streams consist of a fresh fuel\slash air mixture stream at a specified temperature and equivalence ratio, and a pilot stream consisting of the adiabatic equilibrium products of the fresh mixture stream, with the mass flow rates of these two streams in a ratio of $0.95 \mathbin{:} 0.05$.
Non-premixed cases consist of three inflow streams: air, fuel, and a pilot consisting of the adiabatic equilibrium products of a stoichiometric fuel\slash air mixture at the same unburned temperature as the first two streams; the mass flow rates of these streams occur in a ratio of $0.85 \mathbin{:} 0.05 \mathbin{:} 0.1$.
Following inflow\slash outflow (for both premixed and non-premixed cases), $\frac{1}{2} N_{\text{part}} \Delta t / \tau_{\text{pair}}$ pairs of particles (not including the inflowing particles) are randomly selected for pairing and then randomly shuffled with the inflowing particles to exchange partners, where $\tau_{\text{pair}}$ is the pairing timescale.

Although multiple mixing models exist~\cite{Ren:2004fz}, the current mixing fractional step consists of a pair of particles $\phi^p$ and $\phi^q$ exchanging compositional information and evolving by
\begin{align}
\frac{d \phi^p}{dt} &= - \frac{ \phi^p - \phi^q }{ \tau_{\text{mix}} } \quad \text{and} \\
\frac{d \phi^q}{dt} &= - \frac{ \phi^q - \phi^p }{ \tau_{\text{mix}} } \;,
\end{align}
where $\tau_{\text{mix}}$ is a characteristic mixing timescale.
The analytical solution to this system of equations determines the particle compositions $\phi^p$ and $\phi^q$ after a mixture fractional step:
\begin{align}
\phi^p &= \phi^p_0 - \alpha \;, \\
\phi^q &= \phi^q_0 + \alpha \;, \text{ and} \\
\alpha &= \frac{ \phi^p_0 - \phi^q_0 }{2} \left[1 - \exp \left( \frac{-2 \Delta t_{\text{sub}}}{\tau_{\text{mix}}} \right) \right] \;,
\end{align}
where $\phi^p_0$ and $\phi^q_0$ are the particle compositions at the beginning of the mixture fractional step and $\Delta t_{\text{sub}}$ is the mixing sub-time-step size.
The reaction fractional step consists of the enthalpy evolving by
\begin{equation}
\frac{dh}{dt} = \frac{-1}{\rho} \sum_{k=1}^{\numsp} h_k W_k \dot{\omega}_k
\end{equation}
and the species mass fractions evolving according to Eq.~\eqref{e:dTdt}.
However, in practice our implementation handles the reaction fractional step by advancing in time a Cantera~\cite{Goodwin:2015aa} \texttt{ReactorNet} that contains a \texttt{IdealGasConstPressureReactor} object, rather than integrating the above equations directly.

The time integration scheme implemented in our approach determines the discrete time step between inflow\slash outflow and pairing events $\Delta t$ and the sub-time step size $\Delta t_{\text{sub}}$ separating mixing\slash reaction fractional steps, both held constant in the current implementation, via
\begin{align}
\Delta t &= 0.1 \, \min \left( \tau_{\text{res}} , \tau_{\text{pair}} \right) \; \text{and} \\
\Delta t_{\text{sub}} &= 0.04 \, \tau_{\text{mix}} \;,
\end{align}
adopted from Pope~\cite{Pope:1997wu}.

\begin{figure}[tbp]
\centering
\includegraphics[width=.75\textwidth]{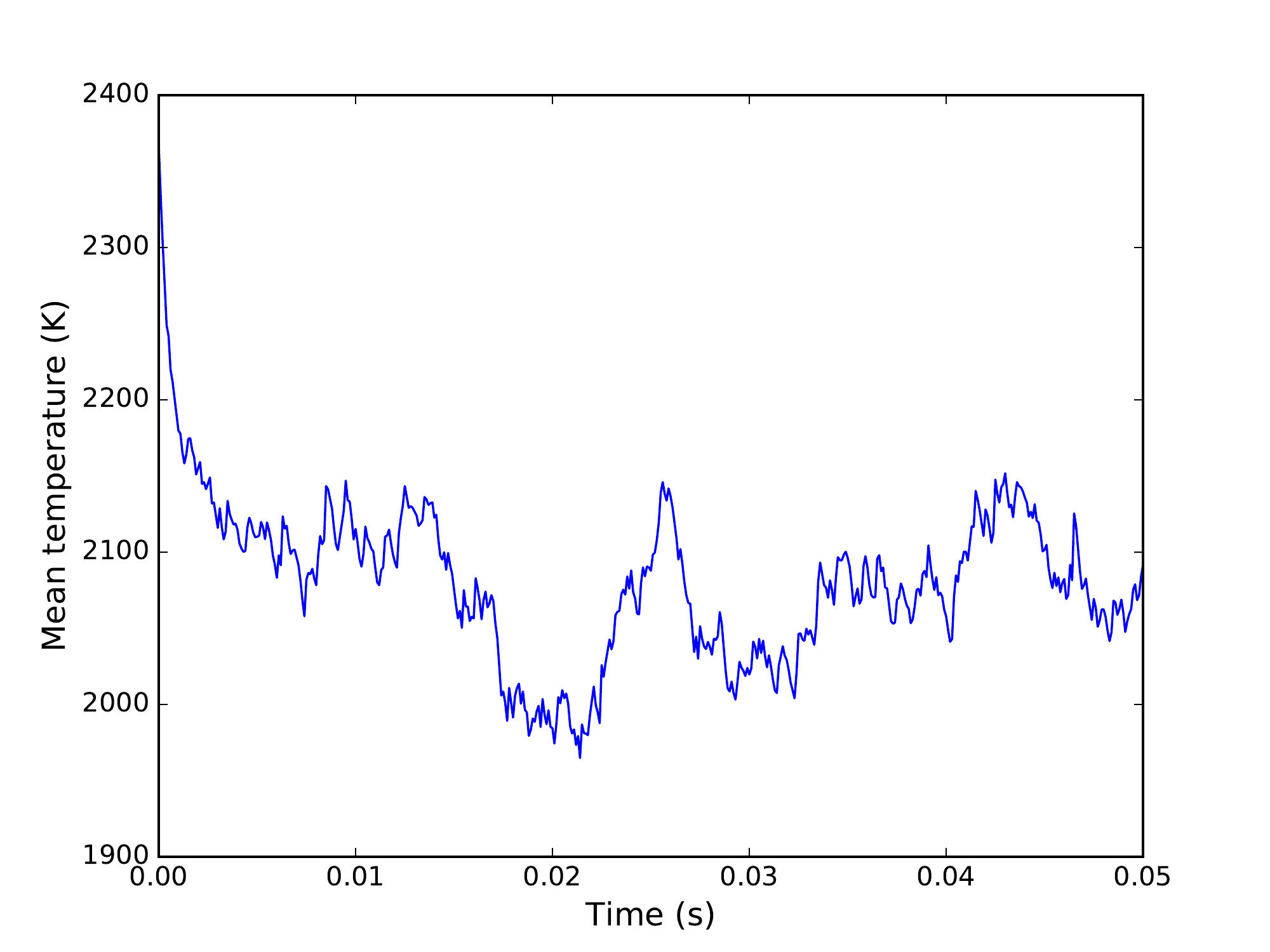}
\caption{Mean temperature of premixed PaSR combustion for stoichiometric
         methane\slash air with an unburned temperature of \SI{600}{\kelvin}
         and at \SI{1}{\atm}, $\tau_{\text{res}} = \SI{5}{\milli\second}$,
         $\tau_{\text{mix}} = \tau_{\text{pair}} = \SI{1}{\milli\second}$,
         and using 100 particles.
         Data, plotting scripts, and the figure file are available under
         CC-BY~\cite{paperscript:2017}.
         }
\label{F:ch4_meantemp}
\end{figure}

\begin{figure}[tbp]
\centering
\includegraphics[width=.75\textwidth]{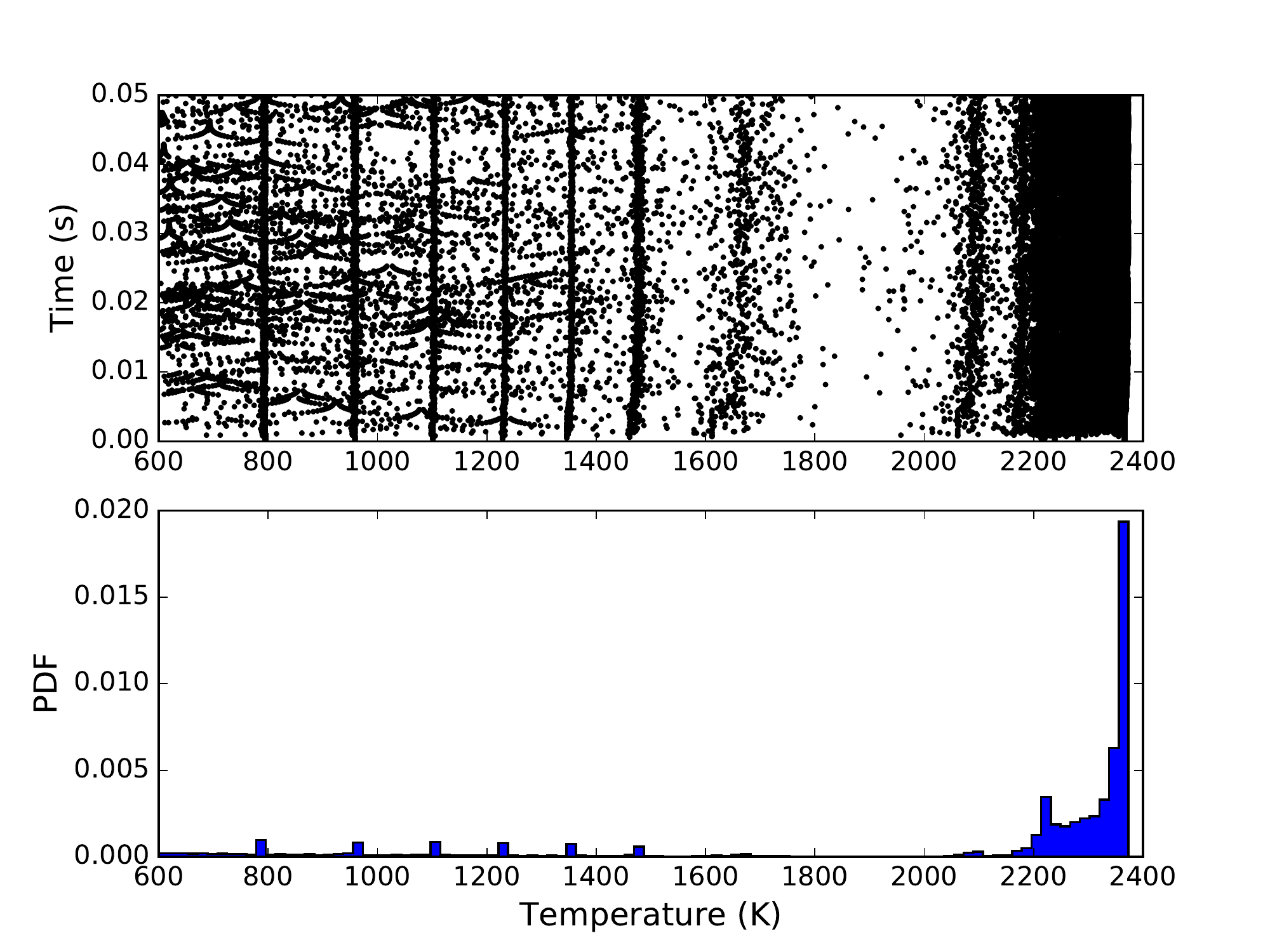}
\caption{Scatterplot of temperature over time (top) and probability density
         function (PDF) of temperature (bottom) of premixed PaSR combustion for
         stoichiometric methane\slash air with an unburned temperature of
         \SI{600}{\kelvin} and at \SI{1}{\atm}, $\tau_{\text{res}} = \SI{5}{\milli\second}$,
         $\tau_{\text{mix}} = \tau_{\text{pair}} = \SI{1}{\milli\second}$,
         and using 100 particles. Data, plotting scripts, and the figure file
         are available under CC-BY~\cite{paperscript:2017}.
         }
\label{F:ch4_particle_temp}
\end{figure}

Figures~\ref{F:ch4_meantemp} and \ref{F:ch4_particle_temp} demonstrate sample results from premixed PaSR combustion of methane\slash air, using GRI-Mech 3.0~\cite{smith_gri-mech_30}; Fig.~\ref{F:ch4_meantemp} shows the mean temperature evolution over time, while Fig.~\ref{F:ch4_particle_temp} shows the temperature distribution among all particles.
Although a large number of particles reside near the equilibrium temperature of \SI{2367}{\kelvin}, the wide distribution in particle states is evident.

\section{Discussion of \texttt{TChem}'s thread-safety}
\label{A:tchem}

During testing, it became evident that \texttt{TChem} v0.2~\cite{TChem:v0.2} is not thread-safe for
parallelization via OpenMP. In other words, when using OpenMP to parallelize the evaluation of multiple
Jacobian matrices, the results differ from those obtained by evaluating the same matrices
in serial.

To investigate this observed phenomenon further, we created a straightforward test using the
\ce{H2}\slash \ce{CO} model of Burke et al.~\cite{Burke:2011fh} and a subset of the PaSR data
discussed elsewhere in this paper. In short, the testing program first used \texttt{TChem} to
evaluate the Jacobian matrices in serial (i.e., in a single-threaded manner)
for \num{100100} states sampled from the PaSR data described in Table~\ref{T:pasr_parameters}.
Next, this process was repeated, but with OpenMP parallelization of the outer loop enabled.
To determine whether \texttt{TChem} is thread-safe, the computed Jacobian matrices from
these two approaches were compared.

We found significant error between the Jacobian matrices computed in serial and
with multiple threads for all cases, demonstrating that the current version of
\texttt{TChem} is not thread-safe.
This self-contained test is available openly~\cite{Curtis2017:tchem}.



\bibliography{analytical-jacobian-paper}
\bibliographystyle{elsarticle-num}

\end{document}